\documentclass[prx,twocolumn]{revtex4}% Physical Review X

\usepackage{graphicx}% Include figure files
\usepackage{dcolumn}% Align table columns on decimal point
\usepackage{bm}% bold math
\usepackage{multirow}% multirow
\usepackage{amsmath,amssymb}% gtrless

\begin{document}

%\preprint{APS/123-QED}

\title{Temperature-driven change in band structure reflecting spin-charge separation of Mott and Kondo insulators}

\author{Masanori Kohno}
\email{KOHNO.Masanori@nims.go.jp}
\affiliation{Research Center for Materials Nanoarchitectonics, National Institute for Materials Science, Tsukuba 305-0003, Japan}

\date{\today}

\begin{abstract}
The electronic band structure can change with temperature in Mott and Kondo insulators, 
even without a phase transition. 
Here, to clarify the underlying mechanism, the spectral function at nonzero temperature is studied. 
By considering the selection rules, the spin excited states of Mott and Kondo insulators, 
whose excitation energies are lower than the charge gap, 
are shown to emerge in the electronic spectral function at nonzero temperature, 
exhibiting momentum-shifted magnetic dispersion relations from the band edges, 
as in the case of the doping-driven Mott transition at zero temperature. 
Based on this characteristic, we interpret the numerical results for the temperature-driven change in the band structure 
in the one- and two-dimensional Hubbard models, Hubbard ladder, and one-dimensional periodic Anderson model at half-filling 
obtained using the cluster perturbation theory with the low-temperature Lanczos method. 
This characteristic also explains why the band structures of Mott and Kondo insulators can change with temperature 
even in the energy regime far higher than the temperature 
and why spectral weights emerge in the energy regime within the band gap, where the excitation energies are lower than 
the lowest electronic excitation energy from the ground state. 
Furthermore, if the bandwidth of the spin excitation is comparable to the electronic band gap, 
the emergent electronic modes can cross the Fermi level and 
gain spectral weight as the temperature increases, which leads to an insulator-metal crossover. 
These features are primarily caused by the spin excited states 
that are transparent in electronic measurements at zero temperature, 
in contrast to the conventional view where thermal effects on electron-added states, electron-removed states, 
and static spin correlations are considered to mainly affect the band structure. 
This innovative perspective provides a different understanding from the conventional view 
on the electronic states at nonzero temperatures. 
\end{abstract}

\maketitle
\section{Introduction} % ------ Introduction ------
Metallic states at high temperatures, where the electronic modes cross the Fermi level, can change to 
insulating states as the temperature decreases in Mott and Kondo insulators. 
An explanation based on the antiferromagnetic mean-field approximation is that an electronic band splits into two 
by opening an energy gap at the Fermi level when an antiferromagnetic phase transition occurs \cite{SlaterAF, PennAF}; 
symmetry breaking is essential for the metal-insulator transition. 
Another explanation based on the dynamical mean-field approximation is that a first-order transition occurs 
between metallic and insulating phases \cite{DMFT_RMP, DMFT_RozenbergPRB}; 
the change from metallic to insulating states is discontinuous. 
\par
However, recent studies on the doping-driven Mott transition at zero temperature 
indicate that the Mott transition can be characterized in terms of the spin-charge separation of Mott and Kondo insulators 
\cite{KohnoRPP,Kohno1DHub,Kohno2DHub,Kohno1DtJ,Kohno2DtJ,KohnoDIS,KohnoAF,KohnoSpin,KohnoHubLadder,KohnoGW,KohnoKLM}, 
i.e., the spin excited states of Mott and Kondo insulators, whose excitation energies are lower than the charge gap, emerge 
as an electronic mode exhibiting a momentum-shifted magnetic dispersion relation upon doping of Mott and Kondo insulators 
\cite{KohnoRPP,Kohno1DHub,Kohno2DHub,Kohno1DtJ,Kohno2DtJ,KohnoDIS,KohnoAF,KohnoSpin,KohnoHubLadder,KohnoGW,KohnoKLM}. 
\par
In this paper, this characteristic is extended to the change in the band structure with respect to the temperature 
in Mott and Kondo insulators by considering the selection rules in the spectral function and 
using numerical methods for the one-dimensional (1D) and two-dimensional (2D) Hubbard models, 
the Hubbard ladder, and the 1D periodic Anderson model (PAM). 
\par
At nonzero temperature, the spin excited states of Mott and Kondo insulators are shown to emerge in the electronic spectral function, 
exhibiting momentum-shifted magnetic dispersion relations from the band edges, as in the case of the doping-driven Mott transition 
\cite{KohnoRPP,Kohno1DHub,Kohno2DHub,Kohno1DtJ,Kohno2DtJ,KohnoDIS,KohnoAF,KohnoSpin,KohnoHubLadder,KohnoGW,KohnoKLM}. 
The emergence of the electronic modes changes the band structure. 
If the emergent electronic modes cross the Fermi level, the band structure can be regarded as metallic. 
This implies that the band structure can change from insulating to metallic as the temperature increases. 
The conditions for this insulator-metal crossover are derived using the selection rules, and 
the temperature evolution of the emergent modes and the insulator-metal crossover are 
demonstrated by numerical calculations. 
This crossover has not been expected in the conventional mean-field approximations 
\cite{SlaterAF, PennAF, DMFT_RMP, DMFT_RozenbergPRB}; 
symmetry breaking is not necessary, and the electronic states continuously evolve from insulating to metallic states 
with an increase in the temperature. 
\par
In the 1D and 2D Hubbard models and 1D PAM, numerical calculations have indicated that 
the band structures change with temperature 
\cite{TPQSLanczos,PreussQP,GroberQMC,KuzminCPT,VCALanczos,NoceraDMRT_T,Matsueda_QMC,PAMMPS}. 
The spectral features have been discussed by comparing the numerical results with the band structures of 
the Hubbard-I approximation \cite{HubbardI} and antiferromagnetic mean-field approximation \cite{SlaterAF, PennAF} 
and interpreted in terms of the static spin correlations \cite{PreussQP,GroberQMC} and 
quasiparticles such as dressed electronic quasiparticles \cite{GroberQMC}, 
spinons, and (anti)holons \cite{NoceraDMRT_T} for electronic excitation. 
\par
In this paper, why and how the band structures of Mott and Kondo insulators change with temperature are clarified 
using the selection rules and numerical calculations. 
The temperature-driven change in the band structure can be primarily explained as resulting from the spin excited states, 
which have not been recognized as crucial in previous studies. 
The reasons why the band structure can change with temperature even in the energy regime far higher 
than the temperature, 
why spectral weights emerge in the energy regime within the band gap, where the excitation energies are lower than 
the lowest electronic excitation energy from the ground state, 
and why the emergent modes exhibit momentum-shifted spin-mode-like dispersion relations from the band edges 
are explained without depending on the spatial dimension, antiferromagnetic order, or quasiparticle picture. 
\par
The emergence of electronic modes due to doping 
\cite{KohnoRPP,Kohno1DHub,Kohno2DHub,Kohno1DtJ,Kohno2DtJ,KohnoDIS,KohnoAF,KohnoSpin,KohnoHubLadder,KohnoGW,KohnoKLM} and temperature is consistently explained in terms of the spin-charge separation of Mott and Kondo insulators 
(existence of spin excited states with excitation energies lower than the charge gap), 
which is a fundamental and general property of Mott and Kondo insulators. 
\section{Model and method} % ------ Model and method ------
The Hubbard model and PAM are defined by the following Hamiltonians: 
\begin{align}
\label{eq:KLM}
{\cal H}_{\rm Hub}&=-\sum_{\langle i,j\rangle,\sigma}t_{i,j}(c^{\dagger}_{i,\sigma}c_{j,\sigma}+{\mbox {H.c.}})\nonumber\\
&+U\sum_{i}\left(n^c_{i,\uparrow}-\frac{1}{2}\right)\left(n^c_{i,\downarrow}-\frac{1}{2}\right)-\mu\sum_{i,\sigma}n^c_{i,\sigma},\\
\label{eq:PAM}
{\cal H}_{\rm PAM}&=-\sum_{\langle i,j\rangle,\sigma}t_{i,j}(c^{\dagger}_{i,\sigma}c_{j,\sigma}+{\mbox {H.c.}})
+U\sum_i n^f_{i,\uparrow}n^f_{i,\downarrow}\nonumber\\
&-t_{\rm K}\sum_{i,\sigma}(c^{\dagger}_{i,\sigma}f_{i,\sigma}+{\mbox {H.c.}})-\Delta\sum_{i,\sigma}n^f_{i,\sigma}\nonumber\\
&-\mu\sum_{i,\sigma}(n^c_{i,\sigma}+n^f_{i,\sigma}), 
\end{align}
where $c_{i,\sigma}$ $(f_{i,\sigma})$ and $n^c_{i,\sigma}$ $(n^f_{i,\sigma})$ denote 
the annihilation and number operators, respectively, of an electron with spin $\sigma$ 
in the conduction (localized) orbital at site $i$. 
Here, $\langle i,j\rangle$ indicates that sites $i$ and $j$ are nearest neighbors; 
$t_{i,j}=t$ in chains and planes; $t_{i,j}=t_\perp$ between chains on a ladder. 
In the PAM, each site has a conduction orbital and localized orbital. 
The numbers of electrons and sites are denoted as $N_{\rm e}$ and $N_{\rm s}$, respectively. 
The thermal average of $N_{\rm e}$ is denoted as ${\bar N}_{\rm e}$. 
The electron density ${\bar n}$ is ${\bar N}_{\rm e}/N$, where $N=N_{\rm s}n_{\rm o}$ 
with $n_{\rm o}$ being the number of orbitals at a site: 
$n_{\rm o}=1$ for the Hubbard model and $2$ for the PAM. 
In this paper, the particle--hole symmetric case at half-filling (${\bar N}_{\rm e}=N$, $\mu=0$ in the Hubbard model; 
${\bar N}_{\rm e}=N$, $\mu=0$, and $U=2\Delta$ in the PAM) for $t,t_{\perp},t_{\rm K},U$, and $\Delta>0$ with even $N$ 
is considered unless otherwise mentioned. 
In the symmetric case, the spectral function is symmetric with respect to $(\omega, {\bm k})=(0,{\bm \pi}/2)$. 
Hereafter, the units $\hbar=1$ and $k_{\rm B}=1$ are used. 
\par
We study the spectral function at temperature $T$, which is defined as 
\begin{align}
\label{eq:AkwT}
A^a({\bm k},\omega)=\frac{1}{2\Xi}\sum_{n,m,\sigma}{\rm e}^{-\beta E_n}
&[|\langle m|a^{\dagger}_{{\bm k},\sigma}|n\rangle|^2\delta(\omega-E_m+E_n)\nonumber\\
+&|\langle m|a_{{\bm k},\sigma}|n\rangle|^2\delta(\omega+E_m-E_n)],
\end{align}
where $\Xi=\sum_{n}{\rm e}^{-\beta E_n}$, $\beta=\frac{1}{T}$, 
and $|n\rangle$ denotes the $n$th eigenstate with energy $E_n$. 
Here, $a_{{\bm k},\sigma}$ represents $c_{{\bm k},\sigma}$ and $f_{{\bm k},\sigma}$; 
$c_{{\bm k},\sigma}$ and $f_{{\bm k},\sigma}$ denote the Fourier transforms 
of $c_{i,\sigma}$ and $f_{i,\sigma}$, respectively. 
\par
The spectral function can also be expressed as 
\begin{equation}
A^a({\bm k},\omega)=-\frac{1}{2\pi}\sum_\sigma{\rm Im}G^a_{{\bm k},\sigma}(\omega)
\end{equation}
through the K{\" a}ll{\'e}n-Lehmann representation of the following Green function: 
\begin{align}
\label{eq:thermalState}
G^a_{{\bm k},\sigma}(\omega)=&-i\int_0^\infty dt {\rm e}^{i\omega t-\epsilon t}\frac{\sum_l{}_l\langle {\rm T}|
\{a_{{\bm k},\sigma}(t),a^\dagger_{{\bm k},\sigma}\}|{\rm T}\rangle_l}{\sum_l{}_l\langle {\rm T}|{\rm T}\rangle_l},\nonumber\\
a_{{\bm k},\sigma}(t)=&{\rm e}^{i{\cal H}t}a_{{\bm k},\sigma}{\rm e}^{-i{\cal H}t},\quad |{\rm T}\rangle_l
={\rm e}^{-\frac{\beta{\cal H}}{2}}|{\rm I}\rangle_l
\end{align}
in the limit of $\epsilon\rightarrow +0$. Here, ${\cal H}$ represents ${\cal H}_{\rm Hub}$ and ${\cal H}_{\rm PAM}$. 
The single-particle density of states is defined as 
\begin{equation}
\label{eq:DOS}
\rho^a(\omega)=\int\frac{d{\bm k}}{(2\pi)^d}A^a({\bm k},\omega)
\end{equation}
in $d$ dimensions for the 1D and 2D Hubbard models ($a=c$) and 1D PAM ($a=c$ and $f$). For the Hubbard ladder, 
\begin{equation}
\label{eq:DOSLad}
\rho^c_{k_y}(\omega)=\int\frac{d k_x}{2\pi}A^c(k_x,k_y,\omega). 
\end{equation}
\par
To investigate the spectral weight in the subspace of specific quantum numbers, 
the spectral function whose thermal-state components, ${\rm e}^{-\frac{\beta E_n}{2}}|n\rangle$ in Eq. (\ref{eq:AkwT}), 
are restricted in the subspace of the number of electrons $N_{\rm e}$ and $z$ component of spin $S^z$ is denoted as 
$A^a({\bm k},\omega)|_{(N_{\rm e},S^z)}$ for $a=c$ and $f$. 
The $N_{\rm e}$ component of $A^a({\bm k},\omega)$ is obtained as 
\begin{equation}
\label{eq:AkwTNe}
A^a({\bm k},\omega)|_{N_{\rm e}}=\sum_{S^z=-S^z_{\rm max}}^{S^z_{\rm max}}A^a({\bm k},\omega)|_{(N_{\rm e},S^z)},
\end{equation}
where $S^z_{\rm max}=\frac{N_{\rm e}}{2}$. 
The $N_{\rm e}$ component of $A^a({\bm k},\omega)$ with spin $S$ is obtained as 
\begin{align}
\label{eq:AkwTS}
A^a({\bm k},\omega)|_{N_{\rm e};S}=\frac{2S+1}{2}\sum_{\eta=\pm 1}
[&A^a({\bm k},\omega)|_{(N_{\rm e},\eta S)}\nonumber\\
-&A^a({\bm k},\omega)|_{(N_{\rm e},\eta(S+1))}], 
\end{align}
where $A^a({\bm k},\omega)|_{(N_{\rm e},\pm(S^z_{\rm max}+1))}=0$. 
The spectral weights of these components are obtained as 
\begin{align}
\label{eq:ATS}
W^a|_{N_{\rm e}}=&\int d\omega\int\frac{d{\bm k}}{(2\pi)^d}A^a({\bm k},\omega)|_{N_{\rm e}},\nonumber\\
W^a|_{N_{\rm e};S}=&\int d\omega\int\frac{d{\bm k}}{(2\pi)^d}A^a({\bm k},\omega)|_{N_{\rm e};S}
\end{align}
in $d$ dimensions for the 1D and 2D Hubbard models ($a=c$) and 1D PAM ($a=c$ and $f$). 
\par
In this paper, the numerical results for the spectral function at nonzero temperatures obtained 
using the cluster perturbation theory (CPT) \cite{CPTPRL,CPTPRB} are shown. 
In the CPT, real-space Green functions $G^a_{i,j,\sigma}(\omega)$ calculated 
with the low-temperature Lanczos method (LTLM) \cite{LTLM} in clusters with $N=12$ were used. 
In the 2D Hubbard model, $(4\times 3)$-site clusters were used; 
the symmetrized spectral function is defined as ${\bar A}^c({\bm k},\omega)=[A^c(k_x,k_y,\omega)+A^c(k_y,k_x,\omega)]/2$. 
In the LTLM, which is equivalent to the thermal-pure-quantum-state method 
with the Lanczos algorithm \cite{TPQS,TPQSLanczos}, 
typically 10 thermal states were generated from random vectors [$|{\rm I}\rangle_l$ in Eq. (\ref{eq:thermalState})] 
in each sector specified by the numbers of up spins $N_{\uparrow}$ and down spins $N_{\downarrow}$. 
The orthonormal states obtained via QR decomposition from the random vectors were used 
as initial block states in the block Lanczos method. 
Typically 600 eigenstates obtained with the block Lanczos method were used as 
$|m\rangle$ and $|n\rangle$ in Eq. (\ref{eq:AkwT}) for each $(N_{\uparrow},N_{\downarrow})$ sector. 
To obtain $G^a_{i,j,\sigma}(\omega)$, $G^a_{i,j,\sigma}(\omega)\Xi$ and $\Xi$ were calculated 
in all the $(N_{\uparrow},N_{\downarrow})$ subspaces separately, 
and the summation was taken over all $(N_{\uparrow},N_{\downarrow})$. 
\par
The numerical results at zero temperature presented in this paper were obtained 
using the non-Abelian dynamical density-matrix renormalization group method 
\cite{KohnoDIS,Kohno1DtJ,Kohno2DtJ,KohnoHubLadder,nonAbelianHub,nonAbeliantJ,nonAbelianThesis,DDMRG} 
under open boundary conditions for $N=120$ (80), where 120 (240) eigenstates of the density matrix were retained 
in the 1D Hubbard model and Hubbard ladder (1D PAM). 
In the 2D Hubbard model, the spectral function at zero temperature was obtained 
using the CPT, where real-space Green functions were calculated with the continued-fraction expansion 
using the simple Lanczos method in $(4\times 4)$-site clusters. 
The transverse dynamical spin susceptibility at zero temperature was obtained 
using the random-phase approximation \cite{KohnoAF, spinbag} 
based on the antiferromagnetic mean-field approximation \cite{SlaterAF, PennAF} in the 2D Hubbard model. 
\par
The dynamical spin structure factors of the Hubbard model and PAM at zero temperature are defined as 
\begin{align}
\label{eq:Skw}
S_{\rm Hub}({\bm k},\omega)&=\sum_{m,\alpha}|\langle m|S^{c,\alpha}_{\bm k}|{\rm GS}\rangle|^2\delta(\omega-e_m),\nonumber\\
S_{\rm PAM}({\bm k},\omega)&=\frac{1}{2}\sum_{m,\alpha}
|\langle m|(S^{c,\alpha}_{\bm k}-S^{f,\alpha}_{\bm k})|{\rm GS}\rangle|^2\delta(\omega-e_m),
\end{align}
where $S^{a,\alpha}_{\bm k}$ denotes the $\alpha$ component of the spin operator ($\alpha=x$, $y$, and $z$) 
with momentum ${\bm k}$ for $a=c$ and $f$, 
and $e_m$ denotes the excitation energy of eigenstate $|m\rangle$ from the ground state $|{\rm GS}\rangle$. 
The transverse dynamical spin susceptibility of the Hubbard model at zero temperature is defined as 
\begin{align}
\label{eq:Skw}
\chi({\bm k},\omega)&=\frac{1}{2}[\chi^{+-}({\bm k},\omega)+\chi^{-+}({\bm k},\omega)],\\
\chi^{\pm\mp}({\bm k},\omega)&=-\sum_{m}
\left[\frac{|\langle m|S^{c,\mp}_{\bm k}|{\rm GS}\rangle|^2}{\omega-e_m+i\epsilon}
-\frac{|\langle m|S^{c,\pm}_{-{\bm k}}|{\rm GS}\rangle|^2}{\omega+e_m+i\epsilon}\right]\nonumber
\end{align}
for $\epsilon\rightarrow +0$, where $S^{c,+}_{\bm k}$ and $S^{c,-}_{\bm k}$ denote the raising and lowering operators, 
respectively, of the $z$ component of spin with momentum ${\bm k}$. 
\section{Spectral features} % ------ Spectral features ------
\label{sec:spectralFeatures}
\begin{figure*}
\includegraphics[width=\linewidth]{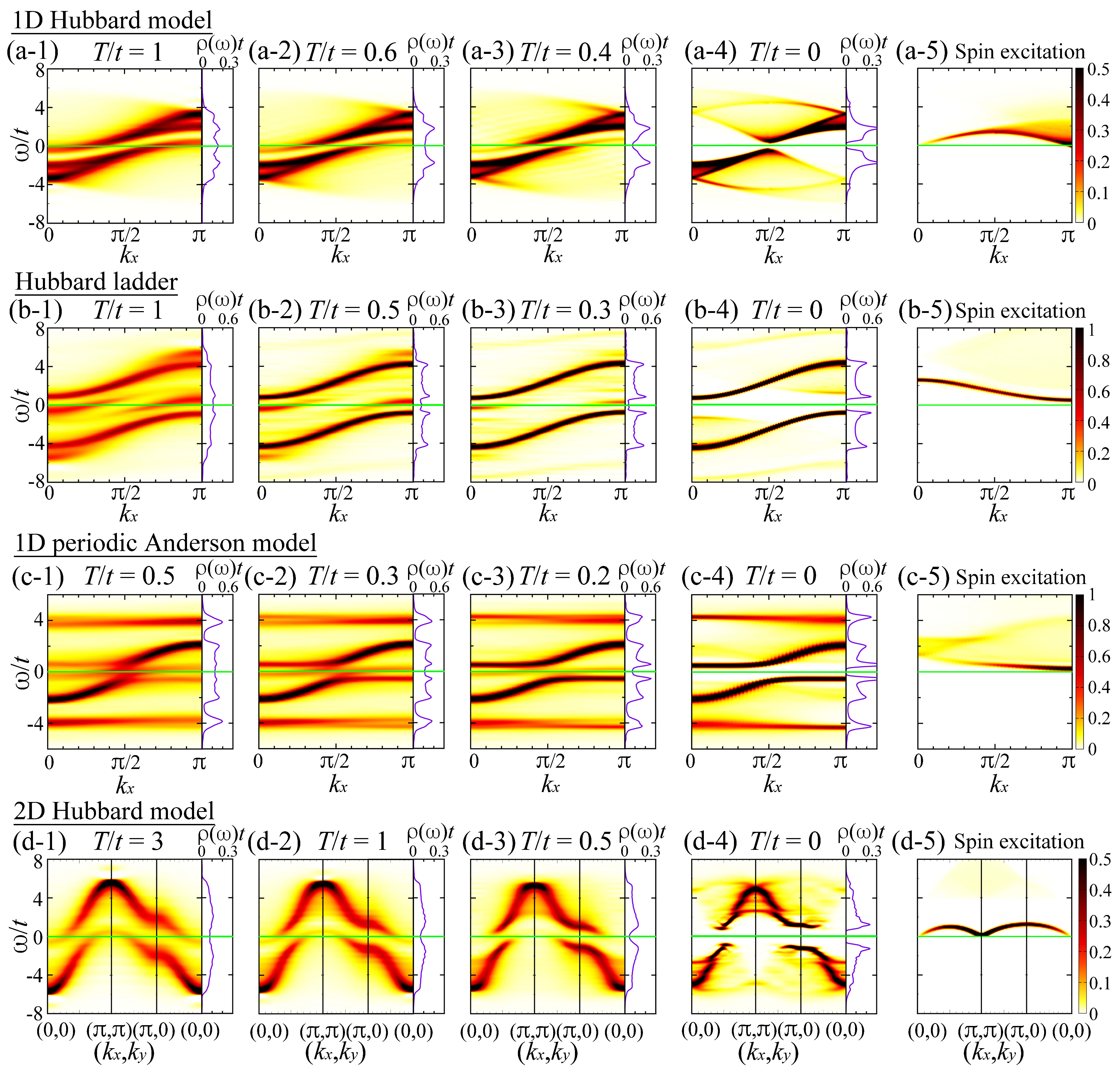}
\caption{Spectral function and spin excitation of the 1D Hubbard model for $U/t=3.4$ [(a-1)--(a-5)], 
Hubbard ladder for $U/t=4$ and $t_\perp/t=2$ [(b-1)--(b-5)], 1D PAM for $U/t=6$ and $t_{\rm K}/t=1.2$ [(c-1)--(c-5)], 
and 2D Hubbard model for $U/t=5$ [(d-1)--(d-5)]. 
(a-1)--(a-4) $A^c(k_x,\omega)t$ at $T/t=1$ [(a-1)], 0.6 [(a-2)], 0.4 [(a-3)], and 0 [(a-4)]. 
(a-5) $S_{\rm Hub}(k_x,\omega)t/3$ at $T=0$. 
(b-1)--(b-4) $A^c(k_x,0,\omega)t+A^c(k_x,\pi,\omega)t$ at $T/t=1$ [(b-1)], 0.5 [(b-2)], 0.3 [(b-3)], and 0 [(b-4)]. 
(b-5) $S_{\rm Hub}(k_x,\pi,\omega)t/3$ at $T=0$. 
(c-1)--(c-4)$A^c(k_x,\omega)t+A^f(k_x,\omega)t$ at $T/t=0.5$ [(c-1)], 0.3 [(c-2)], 0.2 [(c-3)], and 0 [(c-4)]. 
(c-5) $S_{\rm PAM}(k_x,\omega)t/3$ at $T=0$. 
(d-1)--(d-4) ${\bar A}^c({\bm k},\omega)t$ at $T/t=3$ [(d-1)], 1 [(d-2)], 0.5 [(d-3)], and 0 [(d-4)]. 
(d-5) $\frac{1}{2\pi}{\rm Im}\chi({\bm k},\omega)t$ at $T=0$. 
The panels on the right show the single-particle density of states $\rho(\omega)$: $\rho^c(\omega)t$ [(a-1)--(a-4)], 
$\rho^c_{k_y=0}(\omega)t+\rho^c_{k_y=\pi}(\omega)t$ [(b-1)--(b-4)], $\rho^c(\omega)t+\rho^f(\omega)t$ [(c-1)--(c-4)], 
and $\rho^c(\omega)t$ [(d-1)--(d-4)]. 
The green lines indicate $\omega=0$. Gaussian broadening with a standard deviation of $0.1t$ was used.}
\label{fig:Akw}
\end{figure*}
The numerical results for the 1D Hubbard model, Hubbard ladder, 1D PAM, and 2D Hubbard model are shown in Fig. \ref{fig:Akw}. 
At zero temperature, there are no electronic states around the Fermi level ($\omega=0$) [Figs. \ref{fig:Akw}(a-4)--\ref{fig:Akw}(d-4)]. 
However, electronic states emerge within the band gap, 
and their spectral weights increase with temperature 
[Figs. \ref{fig:Akw}(a-1)--\ref{fig:Akw}(a-3), \ref{fig:Akw}(b-1)--\ref{fig:Akw}(b-3), \ref{fig:Akw}(c-1)--\ref{fig:Akw}(c-3), and \ref{fig:Akw}(d-1)--\ref{fig:Akw}(d-3)]. 
When the bandwidth of the spin excitation is comparable to the electronic band gap, as shown in Fig. \ref{fig:Akw}, 
the emergent electronic states form modes crossing the Fermi level, 
which can be regarded as a metallic band structure. 
In this paper, the characteristic modes carrying significant spectral weights are identified as bands. 
\par
This insulator-metal crossover has not been expected in the conventional band picture or 
mean-field approximations \cite{SlaterAF, PennAF, DMFT_RMP, DMFT_RozenbergPRB}. 
The origin and properties of the emergent modes are explained in the following sections 
in terms of the spin-charge separation of Mott and Kondo insulators. 
\section{Emergent electronic modes at nonzero temperature} % ------ Emergent electronic modes at nonzero temperature ------
\label{sec:emergentModes}
\subsection{Quantum numbers} % ------ Quantum numbers ------
\label{sec:quantumNumbers}
To identify the origin of the emergent modes, we consider quantum numbers of states in the spectral function. 
Hereinafter, the number of electrons $N_{\rm e}$, $z$ component of spin $S^z$, 
and momentum ${\bm k}$ of a state are indicated as $(N_{\rm e},S^z,{\bm k})$. 
The $z$ component of spin of an electron is denoted 
as $s^z$: $s^z=\frac{1}{2}$ and $-\frac{1}{2}$ for $\sigma=\uparrow$ and $\downarrow$, respectively. 
The ground state and its energy in the subspace of $N_{\rm e}$ electrons are denoted as 
$|{\rm GS}\rangle_{N_{\rm e}}$ and $E^{\rm GS}_{N_{\rm e}}$, respectively. 
The $\omega$ values at the bottom of the upper band and top of the lower band at zero temperature 
are denoted as $\mu_+$ and $\mu_-$, respectively: 
\begin{equation}
\label{eq:mu}
\begin{split}
\mu_+&=E^{\rm GS}_{N+1}-E^{\rm GS}_{N}(>0),\\
\mu_-&=E^{\rm GS}_{N}-E^{\rm GS}_{N-1}(<0). 
\end{split}
\end{equation}
\par
The spin excited state in the subspace of $N_{\rm e}$ electrons is denoted 
as $|{\rm Spin}\rangle_{N_{\rm e}}$. 
In the spin excitation [Figs. \ref{fig:Akw}(a-5)--\ref{fig:Akw}(d-5)], the spin modes 
(the magnon in antiferromagnetically ordered systems [Fig. \ref{fig:Akw}(d-5)], 
the triplon in spin gap systems [Figs. \ref{fig:Akw}(b-5) and \ref{fig:Akw}(c-5)], and 
the lower edge of the continuum in chains [Fig. \ref{fig:Akw}(a-5)]) are dominant. 
Hence, $|{\rm Spin}\rangle_{N_{\rm e}}$ basically represents the eigenstate of the spin mode 
in the subspace of $N_{\rm e}$ electrons in this paper. 
\par
For brevity, $|{\rm GS}\rangle_{N_{\rm e}}$ and $|{\rm Spin}\rangle_{N_{\rm e}}$ with $(N_{\rm e},S^z,{\bm k})$ 
are denoted as $|{\rm GS}\rangle_{N_{\rm e}}^{S^z,{\bm k}}$ and $|{\rm Spin}\rangle_{N_{\rm e}}^{S^z,{\bm k}}$, respectively. 
The $z$ component of spin of $|{\rm GS}\rangle_{N_{\rm e}}$ with odd $N_{\rm e}$ is denoted as $\zeta$: 
$\zeta=\frac{1}{2}$ or $-\frac{1}{2}$. 
The $z$ component of spin of $|{\rm Spin}\rangle_{N_{\rm e}}$ with even $N_{\rm e}$ is denoted as $s^z_{\rm T}$: 
$s^z_{\rm T}=1$, $0$, or $-1$. 
\par
If $|n\rangle$ has $(N_{\rm e},S^z,{\bm k})$, 
$a^{\dagger}_{{\bm k}^\prime,\sigma}|n\rangle$ has $(N_{\rm e}+1,S^z+s^z,{\bm k}+{\bm k}^\prime)$, and 
$a_{{\bm k}^\prime,\sigma}|n\rangle$ has $(N_{\rm e}-1,S^z-s^z,{\bm k}-{\bm k}^\prime)$. 
Hence, the matrix elements $\langle m|a^{\dagger}_{{\bm k}^\prime,\sigma}|n\rangle$ and $\langle m|a_{{\bm k}^\prime,\sigma}|n\rangle$ 
can be nonzero only when $|m\rangle$ has $(N_{\rm e}+1,S^z+s^z,{\bm k}+{\bm k}^\prime)$ and $(N_{\rm e}-1,S^z-s^z,{\bm k}-{\bm k}^\prime)$, 
respectively. 
\subsection{Zero temperature} % ------ Zero temperature ------
At zero temperature, $|n\rangle$ in Eq. (\ref{eq:AkwT}) is the ground state at half-filling $|{\rm GS}\rangle_{N}^{0,{\bm 0}}$. 
The electron-addition matrix element 
$\langle m|a^{\dagger}_{{\bm k},\sigma}|{\rm GS}\rangle_{N}^{0,{\bm 0}}$ 
and electron-removal matrix element $\langle m|a_{{\bm k},\sigma}|{\rm GS}\rangle_{N}^{0,{\bm 0}}$ can be nonzero 
if $|m\rangle$ has $(N+1,s^z,{\bm k})$ and $(N-1,-s^z,-{\bm k})$, respectively. 
\par
\begin{figure}
\includegraphics[width=\linewidth]{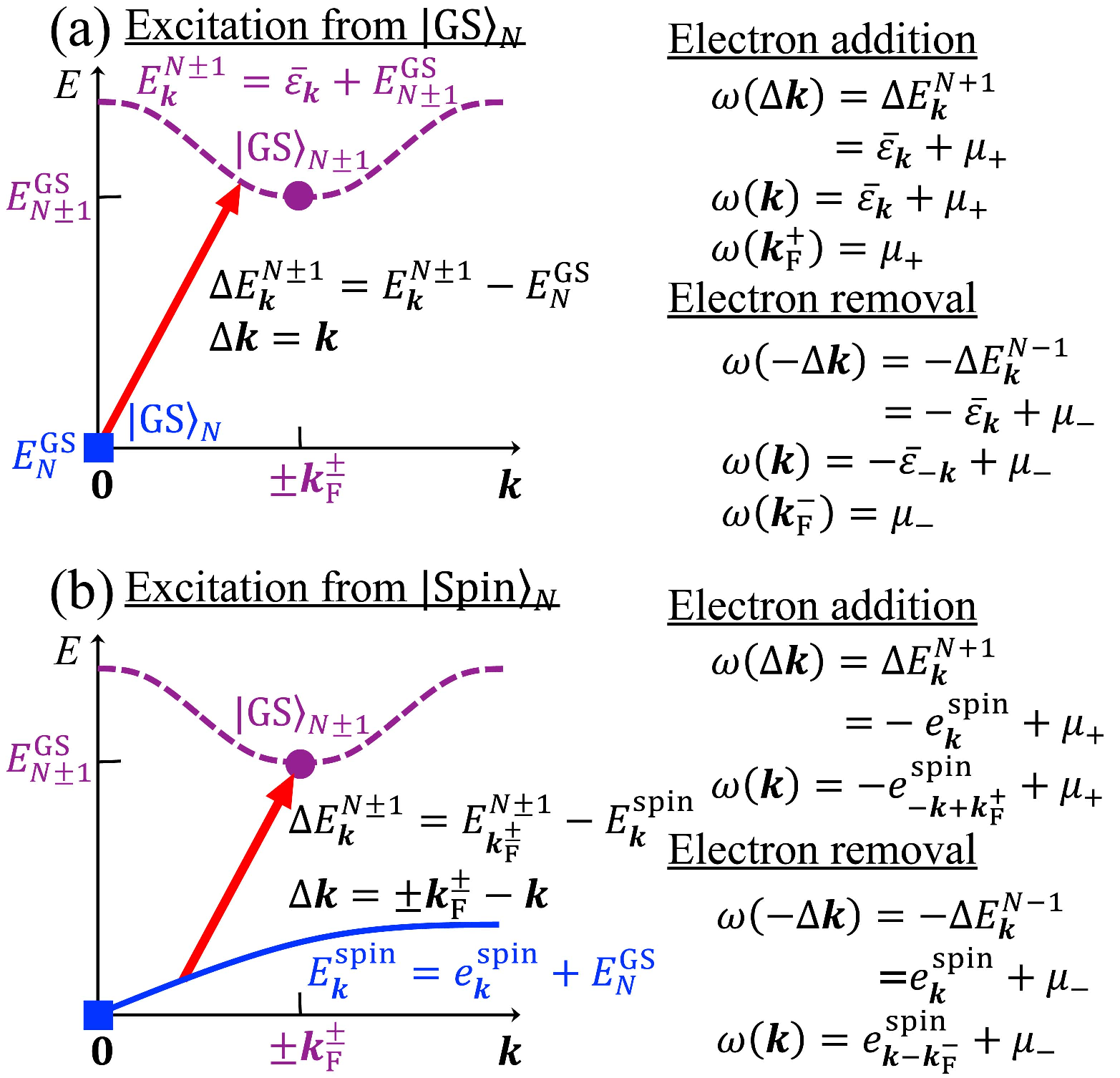}
\caption{Schematic diagrams for the elementary processes of electronic excitation. 
(a) Excitation from the ground state at half-filling $|{\rm GS}\rangle_{N}$ to the one-electron-added (removed) state [$N_{\rm e}=N+1$ ($N-1$)]. 
(b) Excitation from the spin excited state at half-filling $|{\rm Spin}\rangle_{N}$ to the one-electron-added and removed ground states $|{\rm GS}\rangle_{N\pm 1}$. 
The blue square and purple dot represent $|{\rm GS}\rangle_{N}$ and $|{\rm GS}\rangle_{N\pm 1}$, respectively. 
The dashed purple curve represents the energy at momentum ${\bm k}$ of the one-electron-added and removed states, $E_{\bm k}^{N\pm 1}$; 
${\bar \varepsilon}_{\bm k}=E_{\bm k}^{N\pm 1}-E^{\rm GS}_{N\pm 1}$. 
The solid blue curve in (b) represents the energy at momentum ${\bm k}$ of the spin excited state, $E_{\bm k}^{\rm spin}$; 
the excitation energy from $|{\rm GS}\rangle_{N}$ is denoted as $e_{\bm k}^{\rm spin}$. 
The energy and momentum transfers are denoted as $\Delta E_{\bm k}^{N\pm1}$ and $\Delta {\bm k}$, respectively. 
The dispersion relation in the electronic spectrum is denoted as $\omega({\bm k})$.}
\label{fig:ExcitationCartoon}
\end{figure}

The bottom of the upper band at (${\bm k},\omega)=({\bm k}^+_{\rm F},\mu_+)$ corresponds to 
$|m\rangle=|{\rm GS}\rangle_{N+1}^{s^z,{\bm k}^+_{\rm F}}$, and 
the top of the lower band at $({\bm k},\omega)=({\bm k}^-_{\rm F},\mu_-)$ corresponds to 
$|m\rangle=|{\rm GS}\rangle_{N-1}^{-s^z,-{\bm k}^-_{\rm F}}$, 
where ${\bm k}^+_{\rm F}$ and ${\bm k}^-_{\rm F}$ denote the Fermi momenta of the one-electron-doped 
and one-hole-doped systems, respectively [Table \ref{tbl:selectionRule}; Fig. \ref{fig:ExcitationCartoon}(a)]. 
There are no electronic excited states within the band gap at zero temperature. 
\begin{table}
\caption{Selection rules for $\langle m|c^{\dagger}_{{\bm k},\sigma}|n\rangle$ and 
$\langle m|f^{\dagger}_{{\bm k},\sigma}|n\rangle$ at half-filling. 
The energy, $z$ component of spin, and momentum are shown in parentheses. 
The $\omega$ values at the bottom of the upper band and top of the lower band at zero temperature 
are denoted as $\mu_+$ and $\mu_-$, respectively: 
$\mu_+=E^{\rm GS}_{N+1}-E^{\rm GS}_{N}$; 
$\mu_-=E^{\rm GS}_{N}-E^{\rm GS}_{N-1}$. 
The energy of spin excitation with momentum ${\bm q}$ at half-filling is denoted as $e^{\rm spin}_{\bm q}$. 
The $z$ component of spin of $|{\rm GS}\rangle_{N\pm1}$ is denoted as $\zeta$: $\zeta=\frac{1}{2}$ or $-\frac{1}{2}$.}
\label{tbl:selectionRule}
\begin{tabular}{ccc}
\hline\hline
$|m\rangle$&$|n\rangle$&$\omega=E_m-E_n$\\\hline % model
$|{\rm GS}\rangle_{N+1}$&$|{\rm GS}\rangle_{N}$&
\multirow{4}{*}{$\omega=\mu_+$}\\
$\left(\begin{array}{c}
E^{\rm GS}_{N+1}\\
s^z\\
{\bm k}(={\bm k}^+_{\rm F})
\end{array}\right)$&
$\left(\begin{array}{c}
E^{\rm GS}_{N}\\
0\\
{\bm 0}
\end{array}\right)$&\\\hline
$|{\rm GS}\rangle_{N+1}$&$|{\rm Spin}\rangle_N$&
\multirow{4}{*}{$\omega=-e^{\rm spin}_{-{\bm k}+{\bm k}^+_{\rm F}}+\mu_+$}\\
$\left(\begin{array}{c}
E^{\rm GS}_{N+1}\\
\zeta\\
{\bm k}^+_{\rm F}
\end{array}\right)$&
$\left(\begin{array}{c}
e^{\rm spin}_{-{\bm k}+{\bm k}^+_{\rm F}}+E^{\rm GS}_{N}\\
-s^z+\zeta\\
-{\bm k}+{\bm k}^+_{\rm F}
\end{array}\right)$&\\\hline
$|{\rm GS}\rangle_{N}$&$|{\rm GS}\rangle_{N-1}$&
\multirow{4}{*}{$\omega=\mu_-$}\\
$\left(\begin{array}{c}
E^{\rm GS}_{N}\\
0\\
{\bm 0}
\end{array}\right)$&
$\left(\begin{array}{c}
E^{\rm GS}_{N-1}\\
-s^z\\
-{\bm k}(=-{\bm k}^-_{\rm F})
\end{array}\right)$&\\\hline
$|{\rm Spin}\rangle_N$&$|{\rm GS}\rangle_{N-1}$&
\multirow{4}{*}{$\omega=e^{\rm spin}_{{\bm k}-{\bm k}^-_{\rm F}}+\mu_-$}\\
$\left(\begin{array}{c}
e^{\rm spin}_{{\bm k}-{\bm k}^-_{\rm F}}+E^{\rm GS}_{N}\\
s^z+\zeta\\
{\bm k}-{\bm k}^-_{\rm F}
\end{array}\right)$&
$\left(\begin{array}{c}
E^{\rm GS}_{N-1}\\
\zeta\\
-{\bm k}^-_{\rm F}
\end{array}\right)$&\\\hline\hline
\end{tabular}
\end{table}
\subsection{Excitation between spin excited states and one-hole-doped ground state} % ------ Excitation between spin excited states and one-hole-doped ground state ------
\label{sec:holeDope}
At nonzero temperatures, $|n\rangle$ in Eq. (\ref{eq:AkwT}) can be any eigenstate. 
However, because of ${\rm e}^{-\beta E_n}$ in Eq. (\ref{eq:AkwT}), 
low-energy eigenstates primarily contribute to the spectral weight at low temperatures. 
\par
Here, we regard the one-hole-doped ground state $|{\rm GS}\rangle_{N-1}^{\zeta,-{\bm k}^-_{\rm F}}$ 
as $|n\rangle$ in Eq. (\ref{eq:AkwT}). 
In this case, $|m\rangle$ with $(N,s^z+\zeta,{\bm k}-{\bm k}^-_{\rm F})$ can have 
a nonzero electron-addition matrix element 
$\langle m|a^{\dagger}_{{\bm k},\sigma}|{\rm GS}\rangle_{N-1}^{\zeta,-{\bm k}^-_{\rm F}}$. 
\par
The ground state at half-filling $|{\rm GS}\rangle_{N}^{0,{\bm 0}}$ 
can have a nonzero electron-addition matrix element 
when $s^z=-\zeta$ and ${\bm k}={\bm k}^-_{\rm F}$ (Table \ref{tbl:selectionRule}). 
This process yields spectral weight at the top of the lower band: 
${\bm k}={\bm k}^-_{\rm F}$ and $\omega=E^{\rm GS}_{N}-E^{\rm GS}_{N-1}(=\mu_-)$ [Eq. (\ref{eq:mu})]. 
\par
The spin excited state at half-filling $|{\rm Spin}\rangle_N^{s^z+\zeta,{\bm k}-{\bm k}^-_{\rm F}}$ 
can also have a nonzero electron-addition matrix element. 
The spectral weight appears along 
$\omega=e^{\rm spin}_{{\bm k}-{\bm k}^-_{\rm F}}+E^{\rm GS}_{N}-E^{\rm GS}_{N-1}$, 
where the excitation energy of $|{\rm Spin}\rangle_N^{s^z_{\rm T},{\bm q}}$ from the ground state at half-filling 
$|{\rm GS}\rangle_{N}^{0,{\bm 0}}$ is denoted as $e^{\rm spin}_{\bm q}$. 
This implies that an electronic mode emerges, exhibiting the spin-mode dispersion relation shifted 
by the Fermi momentum ${\bm k}^-_{\rm F}$ from the top of the lower band 
(Table \ref{tbl:selectionRule}): 
\begin{equation}
\label{eq:dispersion1}
\omega=e^{\rm spin}_{{\bm k}-{\bm k}^-_{\rm F}}+\mu_-.
\end{equation}
\par
The above explanation can also be simplified as follows. 
Because the thermal state [Eq. (\ref{eq:thermalState})] involves the one-hole-doped ground state, 
the doping-induced states at zero temperature, which can be identified as spin excited states 
of the Mott and Kondo insulators 
\cite{KohnoRPP,Kohno1DHub,Kohno2DHub,Kohno1DtJ,Kohno2DtJ,KohnoDIS,KohnoAF,KohnoSpin,KohnoHubLadder,KohnoGW,KohnoKLM}, 
appear at nonzero temperatures \cite{KohnoKLM}. 
Thus, the electronic mode induced by temperature can be primarily identified as spin excited states 
of Mott and Kondo insulators, as in the case of the doping-induced states at zero temperature 
\cite{KohnoRPP,Kohno1DHub,Kohno2DHub,Kohno1DtJ,Kohno2DtJ,KohnoDIS,KohnoAF,KohnoSpin,KohnoHubLadder,KohnoGW,KohnoKLM}. 
\par
The inverse process is the electronic excitation from the spin excited state at half-filling $|{\rm Spin}\rangle_N$ 
to the one-hole-doped ground state $|{\rm GS}\rangle_{N-1}$ [Fig. \ref{fig:ExcitationCartoon}(b)]. 
If the energy of $|{\rm Spin}\rangle_N$ is lower than the energies of electronic excited states with $N_{\rm e}=N\pm 1$, 
$|{\rm Spin}\rangle_N$ has a larger Boltzmann weight and can occupy a substantial part of the thermal state; 
the electronic excitation from $|{\rm Spin}\rangle_N$ is expected to contribute significantly to the spectral function 
at nonzero temperatures. 
If we regard $|{\rm Spin}\rangle_N^{s^z_{\rm T},{\bm q}}$ as $|n\rangle$ in Eq. (\ref{eq:AkwT}), 
$|m\rangle$ with $(N+1,s^z+s^z_{\rm T},{\bm k}+{\bm q})$ can have a nonzero electron-addition matrix element 
$\langle m|a^{\dagger}_{{\bm k},\sigma}|{\rm Spin}\rangle_N^{s^z_{\rm T},{\bm q}}$, and 
$|m\rangle$ with $(N-1,-s^z+s^z_{\rm T},-{\bm k}+{\bm q})$ can have a nonzero electron-removal matrix element 
$\langle m|a_{{\bm k},\sigma}|{\rm Spin}\rangle_N^{s^z_{\rm T},{\bm q}}$. 
Hence, the one-hole-doped ground state $|{\rm GS}\rangle_{N-1}^{\zeta,-{\bm k}^-_{\rm F}}$ 
can have a nonzero electron-removal matrix element when $-{\bm k}+{\bm q}=-{\bm k}^-_{\rm F}$ and 
$-s^z+s^z_T=\zeta(=\frac{1}{2}$ or $-\frac{1}{2})$ [Table \ref{tbl:selectionRule}; Fig. \ref{fig:ExcitationCartoon}(b)]. 
The spectral weight appears along $\omega=e^{\rm spin}_{{\bm k}-{\bm k}^-_{\rm F}}+E^{\rm GS}_{N}-E^{\rm GS}_{N-1}$, 
which implies that an electronic mode exhibiting the spin-mode dispersion relation shifted by the Fermi momentum 
${\bm k}^-_{\rm F}$ emerges from the top of the lower band [Eq. (\ref{eq:dispersion1})]. 
\par
The above processes (electron-addition excitation from $|{\rm GS}\rangle_{N-1}$ to $|{\rm Spin}\rangle_N$, 
electron-removal excitation from $|{\rm Spin}\rangle_N$ to $|{\rm GS}\rangle_{N-1}$) 
contribute to the spectral function for the same dispersion relation. In fact, Eq. (\ref{eq:AkwT}) can be rewritten as 
\begin{align}
\label{eq:AkwT2}
A^a({\bm k},\omega)=\frac{1}{2\Xi}\sum_{n,m,\sigma}&({\rm e}^{-\beta E_n}+{\rm e}^{-\beta E_m})\nonumber\\
&\times|\langle m|a^{\dagger}_{{\bm k},\sigma}|n\rangle|^2\delta(\omega-E_m+E_n), 
\end{align}
where $|\langle m|a^{\dagger}_{{\bm k},\sigma}|n\rangle|^2=|\langle n|a_{{\bm k},\sigma}|m\rangle|^2$. 
\subsection{Excitation between spin excited states and one-electron-doped ground state} % ------ Excitation between spin excited states and one-electron-doped ground state ------
\label{sec:electronDope}
The above analyses can be extended to the electronic excitation 
between the one-electron-doped ground state $|{\rm GS}\rangle_{N+1}^{\zeta,{\bm k}^+_{\rm F}}$ and 
$|{\rm Spin}\rangle_N^{-s^z+\zeta,-{\bm k}+{\bm k}^+_{\rm F}}$, 
by regarding them as $|m\rangle$ and $|n\rangle$ in Eq. (\ref{eq:AkwT2}), respectively. 
This excitation exhibits 
$\omega=E^{\rm GS}_{N+1}-(e^{\rm spin}_{-{\bm k}+{\bm k}^+_{\rm F}}+E^{\rm GS}_{N})$, which is 
the spin-excitation dispersion relation inverted for $\omega\leftrightarrow-\omega$ and shifted by the Fermi momentum ${\bm k}^+_{\rm F}$, 
hanging down from the bottom of the upper band [Table \ref{tbl:selectionRule}; Fig. \ref{fig:ExcitationCartoon}(b)]: 
\begin{equation}
\label{eq:dispersion2}
\omega=-e^{\rm spin}_{-{\bm k}+{\bm k}^+_{\rm F}}+\mu_+.
\end{equation}
\par
Thus, the spin excited state at half-filling $|{\rm Spin}\rangle_N$ generally emerges within the band gap of the electronic spectrum, 
exhibiting momentum-shifted spin-excitation dispersion relations from the band edges 
[Eqs. (\ref{eq:dispersion1}) and (\ref{eq:dispersion2})] 
at nonzero temperature because it can have nonzero matrix elements 
with the doped ground states $|{\rm GS}\rangle_{N\pm 1}$ 
($_N\langle{\rm Spin}|a^{\dagger}_{{\bm k},\sigma}|{\rm GS}\rangle_{N-1}\ne 0$ and 
$_N\langle{\rm Spin}|a_{{\bm k},\sigma}|{\rm GS}\rangle_{N+1}\ne 0$) [Eq. (\ref{eq:AkwT2})], 
as in the case of the doping-induced states at zero temperature 
\cite{KohnoRPP,Kohno1DHub,Kohno2DHub,Kohno1DtJ,Kohno2DtJ,KohnoDIS,KohnoAF,KohnoSpin,KohnoHubLadder,KohnoGW,KohnoKLM}. 
Note that the electronic states reflecting spin excitation do not appear at half-filling at zero temperature 
because $_N\langle{\rm Spin}|a^{\dagger}_{{\bm k},\sigma}|{\rm GS}\rangle_{N}= 
{_N\langle{\rm Spin}|a_{{\bm k},\sigma}|{\rm GS}\rangle_{N}}=0$, 
where $|{\rm GS}\rangle_{N}$ is the only component in the thermal state at zero temperature. 
\subsection{$\omega$ and ${\bm k}$ regimes of emergent modes} % ------ $\omega$ and ${\bm k}$ regimes of emergent modes ------
\label{sec:regimes}
According to the above quantum-number analyses, 
the spin excited states at half-filling generally appear in the electronic spectrum at nonzero temperature, 
exhibiting the spin-excitation dispersion relation shifted by the Fermi momentum from the top of the lower band 
[Eq. (\ref{eq:dispersion1})] 
and the inverted spin-excitation dispersion relation shifted by the Fermi momentum from the bottom of the upper band 
[Eq. (\ref{eq:dispersion2})]. 
If the spin excitation is gapless, the emergent electronic modes should be gapless from the band edges; 
if the spin excitation has an energy gap, the emergent electronic modes should also have 
the same energy gap in the electronic spectrum. 
\par
Because the lower (upper) band in the momentum regime inside (outside) the Fermi sea 
essentially remains entirely filled (empty) in $|{\rm GS}\rangle_{N-1}$ ($|{\rm GS}\rangle_{N+1}$), 
the spectral weights emerge from the lower (upper) band primarily in the momentum regime 
outside (inside) the Fermi sea in the Hubbard model, 
as in the case of the doping-induced states at zero temperature \cite{KohnoRPP,Kohno1DHub,Kohno2DHub,Kohno1DtJ,Kohno2DtJ,KohnoDIS,KohnoAF,KohnoSpin,KohnoHubLadder,KohnoGW,KohnottpHub,KohnottpJ}. 
The spectral-weight distributions are symmetric with respect to $(\omega, {\bm k})=(0,{\bm \pi}/2)$ 
in the symmetric Hubbard model and symmetric PAM at half-filling [Figs. \ref{fig:Akw}(a--d-1--4)]. 
\subsection{Conditions for metallic emergent modes} % ------ Conditions for metallic emergent electronic modes ------
\label{sec:conditions}
If the band gap (charge gap) is significantly larger than the spin-excitation energy, the emergent electronic modes 
near the band edges do not reach the center of the band gap; 
an insulator-metal crossover, as shown in Fig. \ref{fig:Akw}, does not occur 
if the Fermi level ($\omega=0$) is located near the center of the band gap. 
According to the above analyses, the emergent modes can cross the Fermi level if the following conditions are satisfied: 
\begin{equation}
\begin{split}
\label{eq:conditions}
e^{\rm spin}_{\rm max}+\mu_->0&\quad\mbox{and}\quad e^{\rm spin}_{\rm min}+\mu_-<0;\\
-e^{\rm spin}_{\rm max}+\mu_+<0&\quad\mbox{and}\quad-e^{\rm spin}_{\rm min}+\mu_+>0, 
\end{split}
\end{equation}
where $e^{\rm spin}_{\rm max}$ and $e^{\rm spin}_{\rm min}$ denote the maximum and minimum values of 
the spin-mode excitation energy, respectively. 
In the case of gapless spin excitation, the emergent mode crosses $\omega=0$ 
if the highest excitation energy of the spin mode ($e^{\rm spin}_{\rm max}$) exceeds 
the depth of the top of the lower band ($|\mu_-|$) or the excitation energy to the bottom of the upper band ($\mu_+$): 
\begin{equation}
\label{eq:conditionGapless}
e^{\rm spin}_{\rm max}>|\mu_-|\quad\mbox{or}\quad\mu_+.
\end{equation}
\par
Equation (\ref{eq:conditions}) [Eq. (\ref{eq:conditionGapless})] implies that the spin excitation $(e^{\rm spin})$ and 
charge excitation ($\mu_{\pm}$) determine whether Mott and Kondo insulators can become metallic 
with an increase in the temperature. 
This characteristic reflects the spin-charge separation of Mott and Kondo insulators (Sec. \ref{sec:spinChargeSeparation}). 
\subsection{Multiple-hole-doped and multiple-electron-doped states} % ------ Multiple-hole-doped and multiple-electron-doped states ------
\label{sec:multi}
Thus far, the contributions of the one-hole-doped and one-electron-doped ground states 
to the spectral function have been considered. 
The analyses can be extended to the cases of multiple-hole-doped and multiple-electron-doped states, 
which are relevant when the electronic band gap is comparable to or smaller than the temperature 
or the Fermi level is located near or within the bands. 
\par
The following discussion is based on the fact that regardless of the number of electrons, 
the matrix element $\langle m|a^{\dagger}_{{\bm k},\sigma}|n\rangle$ 
in the spectral function [Eq. (\ref{eq:AkwT2})] can be nonzero only if $|m\rangle$ has the number of electrons, 
$z$ component of spin, and momentum larger than those of $|n\rangle$ by 1, $s^z$, and ${\bm k}$, respectively (Sec. \ref{sec:quantumNumbers}). 
\begin{table}
\caption{Selection rules for $\langle m|c^{\dagger}_{{\bm k},\sigma}|n\rangle$ and 
$\langle m|f^{\dagger}_{{\bm k},\sigma}|n\rangle$ in doped systems. 
The energy, $z$ component of spin, and momentum are shown in parentheses. 
The energy of spin excitation with momentum ${\bm q}$ in the subspace of $N_{\rm e}\pm 1$ electrons 
is denoted as ${\tilde e}^{\rm spin}_{\bm q}$, 
and ${\tilde \mu}_+=E^{\rm GS}_{N_{\rm e}}-E^{\rm GS}_{N_{\rm e}-1}$ for $N_{\rm e}>N$; 
${\tilde \mu}_-=E^{\rm GS}_{N_{\rm e}+1}-E^{\rm GS}_{N_{\rm e}}$ for $N_{\rm e}<N$. 
The momenta of the ground states with $N_{\rm e}+1$ and $N_{\rm e}-1$ electrons in (a) are denoted as 
${\tilde {\bm k}}_{\rm F}^-$ and $-{\tilde {\bm k}}_{\rm F}^+$, respectively, and those in (b) are assumed to be ${\bm 0}$. 
The $z$ component of spin of $|{\rm GS}\rangle_{N_{\rm e}}$ with odd $N_{\rm e}$ is denoted as $\zeta$: 
$\zeta=\frac{1}{2}$ or $-\frac{1}{2}$.}
\label{tbl:selectionRule2}
\begin{tabular}{l}
(a) Even $N_{\rm e}$\\ % even Ne
\begin{tabular}{ccc}
\hline\hline
$|m\rangle$&$|n\rangle$&$\omega=E_m-E_n$\\\hline
$|{\rm Spin}\rangle_{N_{\rm e}+1}$&$|{\rm GS}\rangle_{N_{\rm e}(<N)}$&
\multirow{4}{*}{$\omega={\tilde e}^{\rm spin}_{{\bm k}-{\tilde {\bm k}}^-_{\rm F}}+{\tilde \mu}_-$}\\
$\left(\begin{array}{c}
{\tilde e}^{\rm spin}_{{\bm k}-{\tilde {\bm k}}^-_{\rm F}}+E^{\rm GS}_{N_{\rm e}+1}\\
s^z\\
{\bm k}
\end{array}\right)$&
$\left(\begin{array}{c}
E^{\rm GS}_{N_{\rm e}}\\
0\\
{\bm 0}
\end{array}\right)$&\\\hline\hline
$|n\rangle$&$|m\rangle$&$\omega=E_m-E_n$\\\hline
$|{\rm Spin}\rangle_{N_{\rm e}-1}$&$|{\rm GS}\rangle_{N_{\rm e}(>N)}$&
\multirow{4}{*}{$\omega=-{\tilde e}^{\rm spin}_{-{\bm k}+{\tilde {\bm k}}^+_{\rm F}}+{\tilde \mu}_+$}\\
$\left(\begin{array}{c}
{\tilde e}^{\rm spin}_{-{\bm k}+{\tilde {\bm k}}^+_{\rm F}}+E^{\rm GS}_{N_{\rm e}-1}\\
-s^z\\
-{\bm k}
\end{array}\right)$&
$\left(\begin{array}{c}
E^{\rm GS}_{N_{\rm e}}\\
0\\
{\bm 0}
\end{array}\right)$&\\\hline\hline
\end{tabular}\\\\
(b) Odd $N_{\rm e}$\\ % odd Ne
\begin{tabular}{ccc}
\hline\hline
$|m\rangle$&$|n\rangle$&$\omega=E_m-E_n$\\\hline
$|{\rm Spin}\rangle_{N_{\rm e}+1}$&$|{\rm GS}\rangle_{N_{\rm e}(<N)}$&
\multirow{4}{*}{$\omega={\tilde e}^{\rm spin}_{{\bm k}-{\tilde {\bm k}}^-_{\rm F}}+{\tilde \mu}_-$}\\
$\left(\begin{array}{c}
{\tilde e}^{\rm spin}_{{\bm k}-{\tilde {\bm k}}^-_{\rm F}}+E^{\rm GS}_{N_{\rm e}+1}\\
s^z+\zeta\\
{\bm k}-{\tilde {\bm k}}^-_{\rm F}
\end{array}\right)$&
$\left(\begin{array}{c}
E^{\rm GS}_{N_{\rm e}}\\
\zeta\\
-{\tilde {\bm k}}^-_{\rm F}
\end{array}\right)$&\\\hline\hline
$|n\rangle$&$|m\rangle$&$\omega=E_m-E_n$\\\hline
$|{\rm Spin}\rangle_{N_{\rm e}-1}$&$|{\rm GS}\rangle_{N_{\rm e}(>N)}$&
\multirow{4}{*}{$\omega=-{\tilde e}^{\rm spin}_{-{\bm k}+{\tilde {\bm k}}^+_{\rm F}}+{\tilde \mu}_+$}\\
$\left(\begin{array}{c}
{\tilde e}^{\rm spin}_{-{\bm k}+{\tilde {\bm k}}^+_{\rm F}}+E^{\rm GS}_{N_{\rm e}-1}\\
-s^z+\zeta\\
-{\bm k}+{\tilde {\bm k}}^+_{\rm F}
\end{array}\right)$&
$\left(\begin{array}{c}
E^{\rm GS}_{N_{\rm e}}\\
\zeta\\
{\tilde {\bm k}}^+_{\rm F}
\end{array}\right)$&\\\hline\hline
\end{tabular}
\end{tabular}
\end{table}
\par
In the case of even $N_{\rm e}$, the selection rules are presented in Table \ref{tbl:selectionRule2}(a). 
For $N_{\rm e}<N$, by adding an electron to $|{\rm GS}\rangle_{N_{\rm e}}^{0,{\bm 0}}$, 
$|{\rm Spin}\rangle_{N_{\rm e}+1}^{s^z, {\bm k}}$ with 
spin-excitation energy ${\tilde e}^{\rm spin}_{{\bm k}-{\tilde {\bm k}}^-_{\rm F}}$ 
from $|{\rm GS}\rangle_{N_{\rm e}+1}^{\zeta, {\tilde {\bm k}}^-_{\rm F}}$ 
can emerge in the electronic spectrum, for which the dispersion relation is 
\begin{equation}
\label{eq:multidoping1}
\omega={\tilde e}^{\rm spin}_{{\bm k}-{\tilde {\bm k}}^-_{\rm F}}+{\tilde \mu}_-
\overset{N_{\rm e}\approx N}{\approx} e^{\rm spin}_{{\bm k}-{\bm k}^-_{\rm F}}+\mu_-,
\end{equation}
where ${\tilde \mu}_-=E^{\rm GS}_{N_{\rm e}+1}-E^{\rm GS}_{N_{\rm e}}$. 
\par
For $N_{\rm e}>N$, by removing an electron from $|{\rm GS}\rangle_{N_{\rm e}}^{0,{\bm 0}}$, 
$|{\rm Spin}\rangle_{N_{\rm e}-1}^{-s^z, -{\bm k}}$ with 
spin-excitation energy ${\tilde e}^{\rm spin}_{-{\bm k}+{\tilde {\bm k}}^+_{\rm F}}$ 
from $|{\rm GS}\rangle_{N_{\rm e}-1}^{\zeta, -{\tilde {\bm k}}^+_{\rm F}}$ 
can emerge in the electronic spectrum, and the corresponding dispersion relation is 
\begin{equation}
\label{eq:multidoping2}
\omega=-{\tilde e}^{\rm spin}_{-{\bm k}+{\tilde {\bm k}}^+_{\rm F}}+{\tilde \mu}_+
\overset{N_{\rm e}\approx N}{\approx} -e^{\rm spin}_{-{\bm k}+{\bm k}^+_{\rm F}}+\mu_+ 
\end{equation}
where ${\tilde \mu}_+=E^{\rm GS}_{N_{\rm e}}-E^{\rm GS}_{N_{\rm e}-1}$. 
\par
In the case of odd $N_{\rm e}$, the selection rules are presented in Table \ref{tbl:selectionRule2}(b). 
For $N_{\rm e}<N$, by adding an electron to $|{\rm GS}\rangle_{N_{\rm e}}^{\zeta, -{\tilde {\bm k}}^-_{\rm F}}$, 
$|{\rm Spin}\rangle_{N_{\rm e}+1}^{s^z+\zeta, {\bm k}-{\tilde {\bm k}}^-_{\rm F}}$ with 
spin-excitation energy ${\tilde e}^{\rm spin}_{{\bm k}-{\tilde {\bm k}}^-_{\rm F}}$ 
from $|{\rm GS}\rangle_{N_{\rm e}+1}^{0, {\bm 0}}$ can emerge in the electronic spectrum, 
and the dispersion relation is given by Eq. (\ref{eq:multidoping1}). 
\par
For $N_{\rm e}>N$, by removing an electron from $|{\rm GS}\rangle_{N_{\rm e}}^{\zeta, {\tilde {\bm k}}^+_{\rm F}}$, 
$|{\rm Spin}\rangle_{N_{\rm e}-1}^{-s^z+\zeta, -{\bm k}+{\tilde {\bm k}}^+_{\rm F}}$ with 
spin-excitation energy ${\tilde e}^{\rm spin}_{-{\bm k}+{\tilde {\bm k}}^+_{\rm F}}$ 
from $|{\rm GS}\rangle_{N_{\rm e}-1}^{0, {\bm 0}}$ can emerge in the electronic spectrum, 
and the dispersion relation is given by Eq. (\ref{eq:multidoping2}). 
\par
In both even and odd $N_{\rm e}$ cases, 
the emergent electronic modes exhibit the spin-mode dispersion relation shifted by the Fermi momentum. 
This is because the difference in momentum between the ground states with $N_{\rm e}\pm 1$ and $N_{\rm e}$ is 
the Fermi momentum regardless of whether the number of electrons is even or odd. 
The spin-mode dispersion relation near half-filling ($\omega={\tilde e}^{\rm spin}_{\bm q}$) is 
typically almost identical to that at half-filling ($\omega=e^{\rm spin}_{\bm q}$), 
and the Fermi momentum near half-filling (${\tilde {\bm k}}^{\pm}_{\rm F}$) is also 
almost identical to that in the small-doping limit (${\bm k}^{\pm}_{\rm F}$). 
Thus, the emergent electronic modes reflecting the spin excitation 
obtained using multiple-hole-doped and multiple-electron-doped states near half-filling ($N_{\rm e}\approx N$) 
exhibit dispersion relations almost identical to those of the one-hole-doped and one-electron-doped cases, respectively 
[Eqs. (\ref{eq:dispersion1}), (\ref{eq:dispersion2}), (\ref{eq:multidoping1}), and (\ref{eq:multidoping2})] 
\cite{KohnoRPP,Kohno1DHub,Kohno2DHub,Kohno1DtJ,Kohno2DtJ,KohnoDIS,KohnoAF,KohnoSpin,KohnoHubLadder,KohnoGW,KohnoKLM}. 
\par
In a doped system, the spin and charge excitations obtained as particle-hole excitations within a band exist, as in the case of a noninteracting metal. 
Although the selection rules (Table \ref{tbl:selectionRule2}) are also valid for the spin excited states of the particle-hole excitations, 
most of these excited states usually contribute to the almost featureless background behind the characteristic modes. 
\subsection{Doped-state contributions} % ------ Doped-state contributions ------
\begin{figure}
\includegraphics[width=\linewidth]{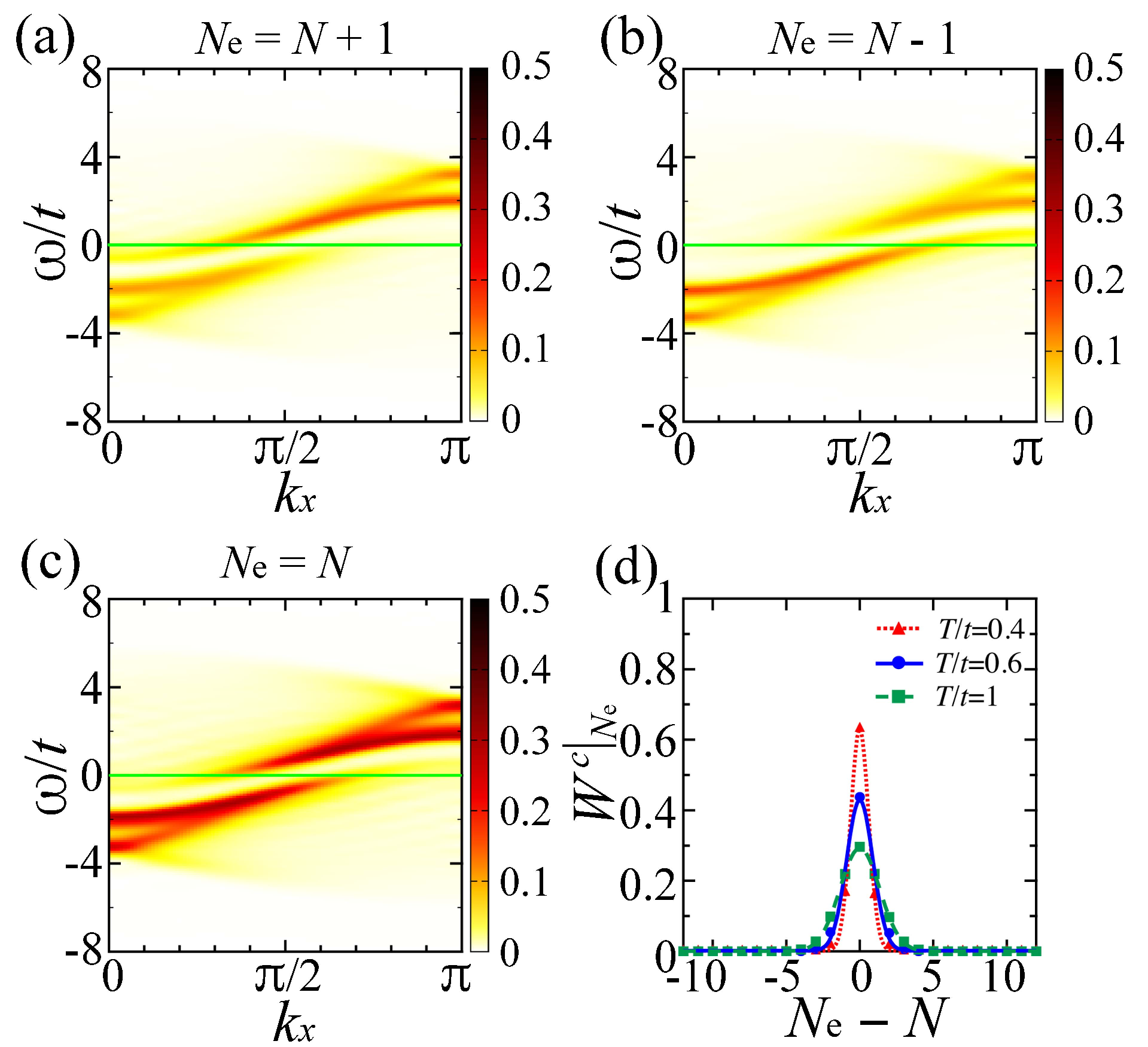}
\caption{Spectral function restricted in the subspace of $N_{\rm e}$ electrons of the thermal state 
in the 1D Hubbard model for $U/t=3.4$ at $T/t=0.6$. 
(a) $A^c(k_x,\omega)|_{N_{\rm e}=N+1}t$. 
(b) $A^c(k_x,\omega)|_{N_{\rm e}=N-1}t$. 
(c) $A^c(k_x,\omega)|_{N_{\rm e}=N}t$. 
(d) $W^c|_{N_{\rm e}}$ at $T/t=0.4$ (red triangles with dotted curve), $T/t=0.6$ (blue circles with solid curve), 
and $T/t=1$ (green squares with dashed curve). The curves are guides for the eye. 
The green lines in (a)--(c) indicate $\omega=0$. Gaussian broadening with a standard deviation of $0.1t$ was used.}
\label{fig:AneU034}
\end{figure}

\label{sec:dopedStates}
When the band gap is comparable to or smaller than the temperature, 
doped components ($N_{\rm e}\ne N$) in the thermal state can contribute substantially to the spectral function 
(Secs. \ref{sec:holeDope}, \ref{sec:electronDope}, and \ref{sec:multi}). 
As in the case of the doping-induced states at zero temperature 
\cite{KohnoRPP,Kohno1DHub,Kohno2DHub,Kohno1DtJ,Kohno2DtJ,KohnoDIS,KohnoAF,KohnoSpin,KohnoHubLadder,KohnoGW,KohnoKLM,NoceraDMRT_T}, 
electron-doped (hole-doped) components in the thermal state yield a considerable spectral weight of the emergent mode 
from the bottom of the upper band (top of the lower band) [Figs. \ref{fig:AneU034}(a) and \ref{fig:AneU034}(b)]. 
\par
As the temperature increases, larger-doping components increase [Fig. \ref{fig:AneU034}(d)] 
because their Boltzmann weights become considerable with the temperature. 
Hence, the intensity of the emergent modes increases with the temperature, 
although the line shape becomes broader because of the accumulation of different-$N_{\rm e}$ modes 
with slightly different dispersion relations (Sec. \ref{sec:multi}) 
as well as contributions from other processes (Sec. \ref{sec:effectiveModes}). 
\par
Note that the undoped component ($N_{\rm e}=N$) in the thermal state also contributes to the emergent modes 
[Fig. \ref{fig:AneU034}(c)] as well as to the upper and lower bands that exist at zero temperature [Fig. \ref{fig:Akw}(a-4)]. 
This is because spin excited states as well as spin-singlet states are included in the undoped component in the thermal state 
(Sec. \ref{sec:reason}). 
\par
\begin{figure}
\includegraphics[width=\linewidth]{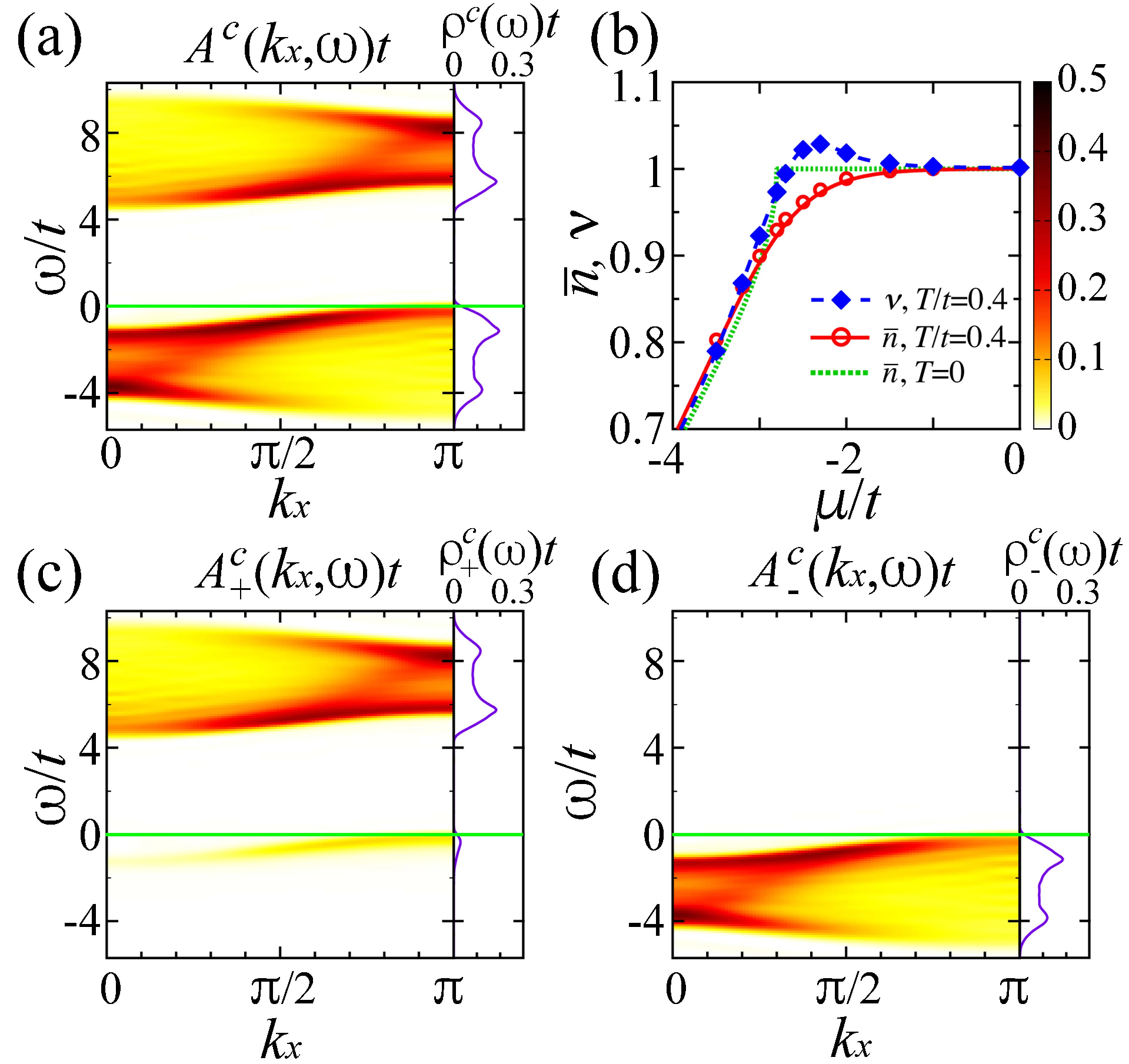}
\caption{Spectral function at $\mu/t=-2.3$, which is slightly above $\mu_-/t\approx -2.8$ for $\mu=0$, 
in the 1D Hubbard model for $U/t=9$ at $T/t=0.4$. 
(a) $A^c(k_x,\omega)t $. (b) Electron density ${\bar n}$ (open red circles) and 
$\nu=2\int_{-\infty}^0d\omega\rho^c(\omega)$ (solid blue diamonds) at $T/t=0.4$. 
The dashed blue line is a cubic spline fit to $\nu$. 
The dotted green and solid red lines indicate ${\bar n}$ obtained using the Bethe ansatz at $T/t=0$ and $0.4$, respectively. 
(c) $A^c_+(k_x,\omega)t$. 
(d) $A^c_-(k_x,\omega)t$. The panels on the right in (a), (c), and (d) show the single-particle density of states: 
$\rho^c(\omega)t$ [(a)], $\rho^c_+(\omega)t$ [(c)], and $\rho^c_-(\omega)t$ [(d)]. 
The green lines in (a), (c), and (d) indicate $\omega=0$. Gaussian broadening with a standard deviation of $0.1t$ was used.}
\label{fig:AneU09}
\end{figure}

Even if the band gap is significantly larger than the temperature, hole-doped components ($N_{\rm e}<N$) 
in the thermal state can contribute considerably to the spectral function 
when the chemical potential is lowered and the Fermi level is located near the top of the lower band (Fig. \ref{fig:AneU09}). 
The increase in hole-doped components in the thermal state enhances 
the spectral weight of the mode emerging from the lower band, 
whereas the decrease in electron-doped components caused by lowering the chemical potential reduces the spectral weight of 
the mode emerging from the upper band. This can be explained as the doping effect 
\cite{KohnoRPP,Kohno1DHub,Kohno2DHub,Kohno1DtJ,Kohno2DtJ,KohnoDIS,KohnoAF,KohnoSpin,KohnoHubLadder,KohnoGW,KohnoKLM,Eskes,DagottoDOS} in the thermal states (Secs. \ref{sec:holeDope} and \ref{sec:electronDope}). 
\par
This feature implies that the spectral weight below the Fermi level can increase 
as the chemical potential is lowered toward the top of the lower band [blue solid diamonds in Fig. \ref{fig:AneU09}(b)]. 
However, this does not mean that the electron density increases as the chemical potential is lowered. 
The increase in the spectral weight is primarily due to the electron-addition excitation from the hole-doped states 
[$\omega<0$ in Fig. \ref{fig:AneU09}(c)], which does not contribute to the electron density. 
In the $d$-dimensional Hubbard model, the spectral weight below the Fermi level is obtained as 
\begin{equation}
\frac{\nu}{2}=\int_{-\infty}^0d\omega \int\frac{d{\bm k}}{(2\pi)^d}A^c({\bm k},\omega), 
\end{equation}
whereas the electron density is obtained as 
\begin{equation}
{\bar n}=2\int_{-\infty}^{\infty}d\omega \int\frac{d{\bm k}}{(2\pi)^d}A^c_-({\bm k},\omega). 
\end{equation}
Here, the spectral functions for electron-addition and electron-removal excitations are defined as follows: 
\begin{align}
\label{eq:AkwTpm}
A^c_+({\bm k},\omega)=\frac{1}{2\Xi}\sum_{n,m,\sigma}{\rm e}^{-\beta E_n}
&|\langle m|a^{\dagger}_{{\bm k},\sigma}|n\rangle|^2\delta(\omega-E_m+E_n),\nonumber\\
A^c_-({\bm k},\omega)=\frac{1}{2\Xi}\sum_{n,m,\sigma}{\rm e}^{-\beta E_n}
&|\langle m|a_{{\bm k},\sigma}|n\rangle|^2\delta(\omega+E_m-E_n), 
\end{align}
where $\Xi=\sum_{n}{\rm e}^{-\beta E_n}$ [Eq. (\ref{eq:AkwT})]. 
The single-particle density of states for $A^c_{\pm}({\bm k},\omega)$ is defined as 
\begin{equation}
\label{eq:DOSAR}
\rho^c_{\pm}(\omega)=\int\frac{d{\bm k}}{(2\pi)^d}A^c_{\pm}({\bm k},\omega).
\end{equation}
\par
As shown in Fig. \ref{fig:AneU09}(b), the electron density decreases monotonically as the chemical potential is lowered 
[open red circles in Fig. \ref{fig:AneU09}(b)], 
which is consistent with the results obtained using the Bethe ansatz [solid red line in Fig. \ref{fig:AneU09}(b)] 
\cite{TakahashiT,TakahashiBook,Essler,LiebWu}, 
whereas the spectral weight below the Fermi level changes nonmonotonically with respect to the chemical potential 
[solid blue diamonds in Fig. \ref{fig:AneU09}(b)]. 
This feature reflects the change in the spectral-weight distribution with respect to the Fermi level. 
In fact, if the spectral-weight distribution is assumed to be unaffected by the Fermi level, 
the spectral weight below the Fermi level decreases monotonically as the chemical potential is lowered. 
This spectral feature cannot be explained in the conventional band picture or 
mean-field approximations (Secs. \ref{sec:AFMF} and \ref{sec:HubI}).
\par
When the Fermi level enters the lower band by lowering the chemical potential further, 
the components of $N_{\rm e}\approx{\bar N}_{\rm e}$ in the thermal state mainly contribute to the spectral function. 
The spectral function is continuously deformed into the zero-temperature spectral function of the doped system 
\cite{KohnoRPP,Kohno1DHub,Kohno2DHub,Kohno1DtJ,Kohno2DtJ,KohnoDIS,KohnoAF,KohnoSpin,KohnoHubLadder,KohnoGW,KohnoKLM} 
with the decreasing of the temperature. 
\subsection{Reason why band structure can change in the energy regime far higher than temperature} % ------ Reason why band structure can change in the energy regime far higher than temperature ------
\label{sec:reason}
In Mott and Kondo insulators whose electronic band gap is significantly larger than the temperature, 
doped states ($N_{\rm e}\ne N$) contribute little to the thermal state 
if the Fermi level is located far from the band edges ($|\mu_{\pm}|\gg T$). 
Nevertheless, Eq. (\ref{eq:AkwT2}) indicates that the spectral weight can emerge 
along the same dispersion relation as the doping-induced states 
if the spin excited states of Mott and Kondo insulators have excitation energies comparable to or lower than the temperature 
because ${\rm e}^{-\beta e^{\rm spin}_{\bm q}}$ can be considerable even if ${\rm e}^{-\beta|\mu_\pm|}$ is negligibly small 
in the spectral function. 
\begin{figure}
\includegraphics[width=\linewidth]{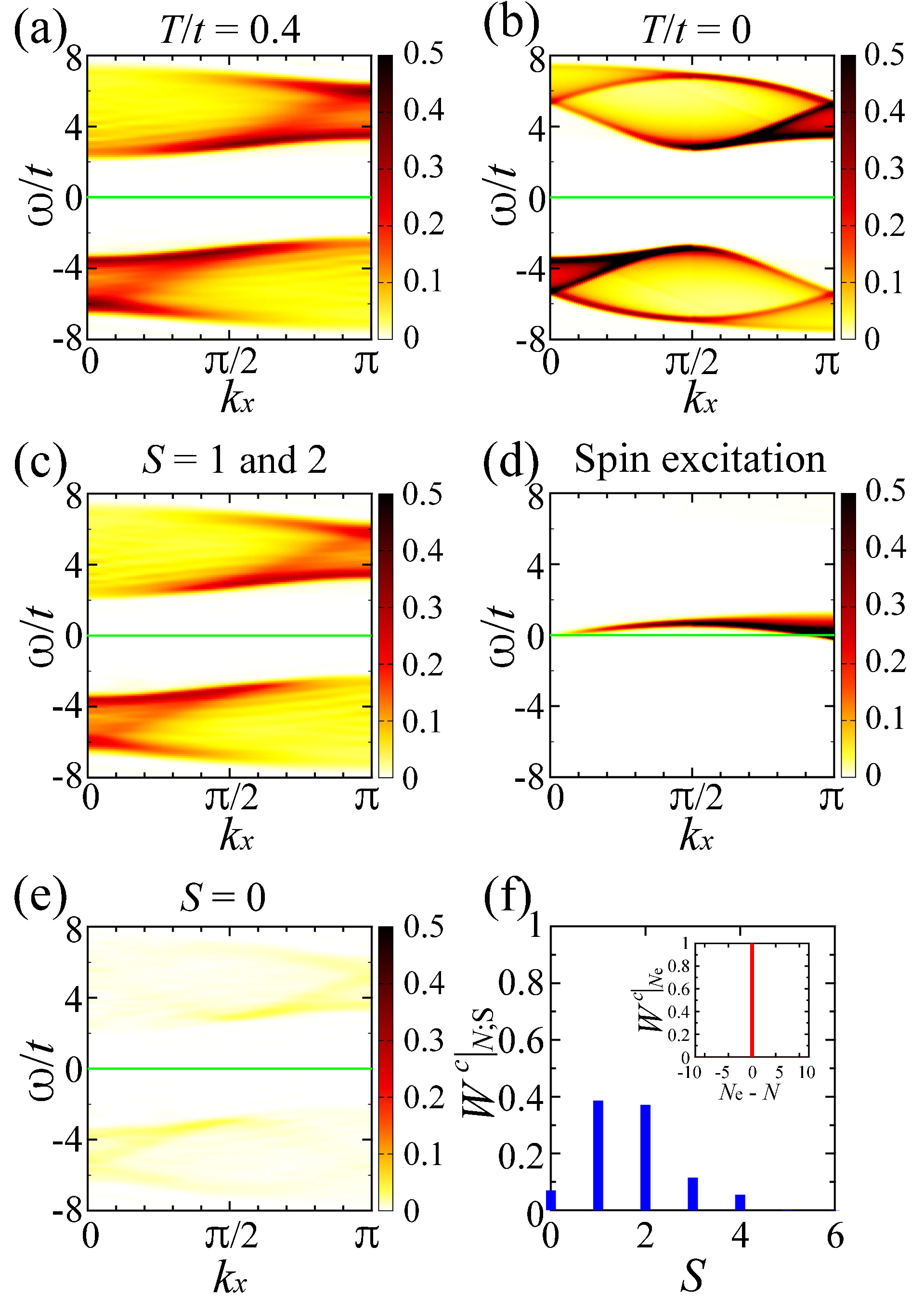}
\caption{Spectral function and spin excitation of the 1D Hubbard model for $U/t=9$. 
(a) $A^c(k_x,\omega)t $ at $T/t=0.4$. (b) $A^c(k_x,\omega)t $ at $T/t=0$. 
(c) $A^c(k_x,\omega)|_{N;S=1}t+A^c(k_x,\omega)|_{N;S=2}t $ at $T/t=0.4$. 
(d) $S_{\rm Hub}(k_x,\omega)t/3$ at $T/t=0$. 
(e) $A^c(k_x,\omega)|_{N;S=0}t $ at $T/t=0.4$. 
(f) $W^c|_{N;S} $ at $T/t=0.4$. The inset shows $W^c|_{N_{\rm e}}$.
The green lines in (a)--(e) indicate $\omega=0$. Gaussian broadening with a standard deviation of $0.1t$ was used.}
\label{fig:AS1D}
\end{figure}
\par
Thus, even if the electronic band gap is significantly larger than the temperature, 
electronic modes can emerge, reflecting the spin excitation in Mott and Kondo insulators [Fig. \ref{fig:ExcitationCartoon}(b)]. 
The spectral weights of the emergent modes increase with temperature 
because the contribution of the spin excited states to the thermal state increases. 
This explains why the band structure changes with temperature in Mott and Kondo insulators 
even in the $|\omega|$ regime far higher than the temperature. 
In other words, the spin-charge separation of Mott and Kondo insulators 
[existence of spin excited states with excitation energies lower than the band gap (charge gap)] 
causes the change in the band structure 
even if the electronic excitation energies to the bands are significantly higher than temperature, 
provided that the temperature is comparable to or higher than the spin-excitation energies. 
\par
Figure \ref{fig:AS1D} shows the change in the band structure for $|\mu_{\pm}|\gg T$ in the 1D Hubbard model. 
The excitation energies from $|{\rm GS}\rangle_N$ to $|{\rm GS}\rangle_{N\pm 1}$ ($|\mu_{\pm}|\approx 2.8t$) 
[band edges in Fig. \ref{fig:AS1D}(b)] are far higher than the temperature ($T=0.4t$). 
The temperature is comparable to the energy scale of the spin excitation $J=\frac{4t^2}{U}\approx 0.4t$ 
[Fig. \ref{fig:AS1D}(d)]. 
Although the bottom of the upper band and top of the lower band at $T=0$ are located at $k_x=\frac{\pi}{2}$ 
[Fig. \ref{fig:AS1D}(b)], those at $T=0.4t$ are located at $k_x=0$ and $\pi$, respectively [Fig. \ref{fig:AS1D}(a)], 
reflecting the spin-mode dispersion relation [Fig. \ref{fig:AS1D}(d)] shifted 
by the Fermi momentum $k^{\pm}_{\rm F}=\frac{\pi}{2}$ [Eqs. (\ref{eq:dispersion1}) and (\ref{eq:dispersion2})]. 
This indicates that the band structure is changed by the temperature which is significantly lower than 
the lowest electronic excitation energy from $|{\rm GS}\rangle_N$ ($T\ll|\mu_{\pm}|$) 
and comparable to the spin-excitation energies ($T\approx J$). 
Although similar results have been obtained by numerical calculations 
\cite{TPQSLanczos,PreussQP,GroberQMC,KuzminCPT,VCALanczos,NoceraDMRT_T,Matsueda_QMC,PAMMPS}, 
the crucial role of the spin excited states of Mott and Kondo insulators has not been recognized (Sec. \ref{sec:comp}). 
\par
At the temperature far lower than the lowest electronic excitation energy from $|{\rm GS}\rangle_N$ 
($T\ll|\mu_{\pm}|$), 
the thermal state basically consists only of $N_{\rm e}=N$ states [inset of Fig. \ref{fig:AS1D}(f)]. 
Among the $N_{\rm e}=N$ states, the spin-triplet ($S=1$) and spin-quintet ($S=2$) states are dominant 
[Figs. \ref{fig:AS1D}(c) and \ref{fig:AS1D}(f)]. 
The spectral weight from the spin-singlet ($S=0$) states, which include the ground state, is significantly smaller than 
that of $S=1$ and $2$ already at $T=0.4t$ [Figs. \ref{fig:AS1D}(e) and \ref{fig:AS1D}(f)]. 
Because the thermal state consists only of the ground state at zero temperature, 
the distribution of $W^c|_{N;S}$ changes from $\delta_{S,0}$ to a distribution that has a maximum at a small but nonzero $S$ 
as the temperature increases [Fig. \ref{fig:AS1D}(f)]. 
Figure \ref{fig:AS1D} indicates that the dominant components in the thermal state for the spectral function 
are the spin excited states with $S>0$ rather than the ground state 
when the temperature is comparable to the spin-excitation energies. 
\subsection{Interpretation of spectral features} % ------ Interpretations of spectral features ------
\label{sec:interpretation}
Based on the selection rules, the spectral features shown in Fig. \ref{fig:Akw} are interpreted below. 
In the low-temperature regime, spectral weights emerge around $\omega=0$ [Figs. \ref{fig:Akw}(a-3)--\ref{fig:Akw}(d-3)] 
within the zero-temperature band gap ($\mu_-<\omega<\mu_+$) [Figs. \ref{fig:Akw}(a-4)--\ref{fig:Akw}(d-4)]. 
According to the selection rules (Table \ref{tbl:selectionRule}), the emergent electronic modes within the band gap 
exhibit the spin-mode dispersion relation shifted by ${\bm k}^{\pm}_{\rm F}$ and $\mu_{\pm}$ 
[Eqs. (\ref{eq:dispersion1}) and (\ref{eq:dispersion2})]. 
The Fermi momentum at the top of the lower band ${\bm k}^-_{\rm F}$ is $\frac{\pi}{2}$ in the 1D Hubbard model, 
$(\pi,0)$ in the Hubbard ladder, $\pi$ in the 1D PAM, and $\frac{\bm \pi}{2}$ in the 2D Hubbard model. 
The Fermi momentum at the bottom of the upper band ${\bm k}^+_{\rm F}$ is $\frac{\pi}{2}$ in the 1D Hubbard model, 
$(0,\pi)$ in the Hubbard ladder, $0$ in the 1D PAM, and $\frac{\bm \pi}{2}$ in the 2D Hubbard model 
[Figs. \ref{fig:Akw}(a-4)--\ref{fig:Akw}(d-4)]. 
\par
\begin{figure}
\includegraphics[width=\linewidth]{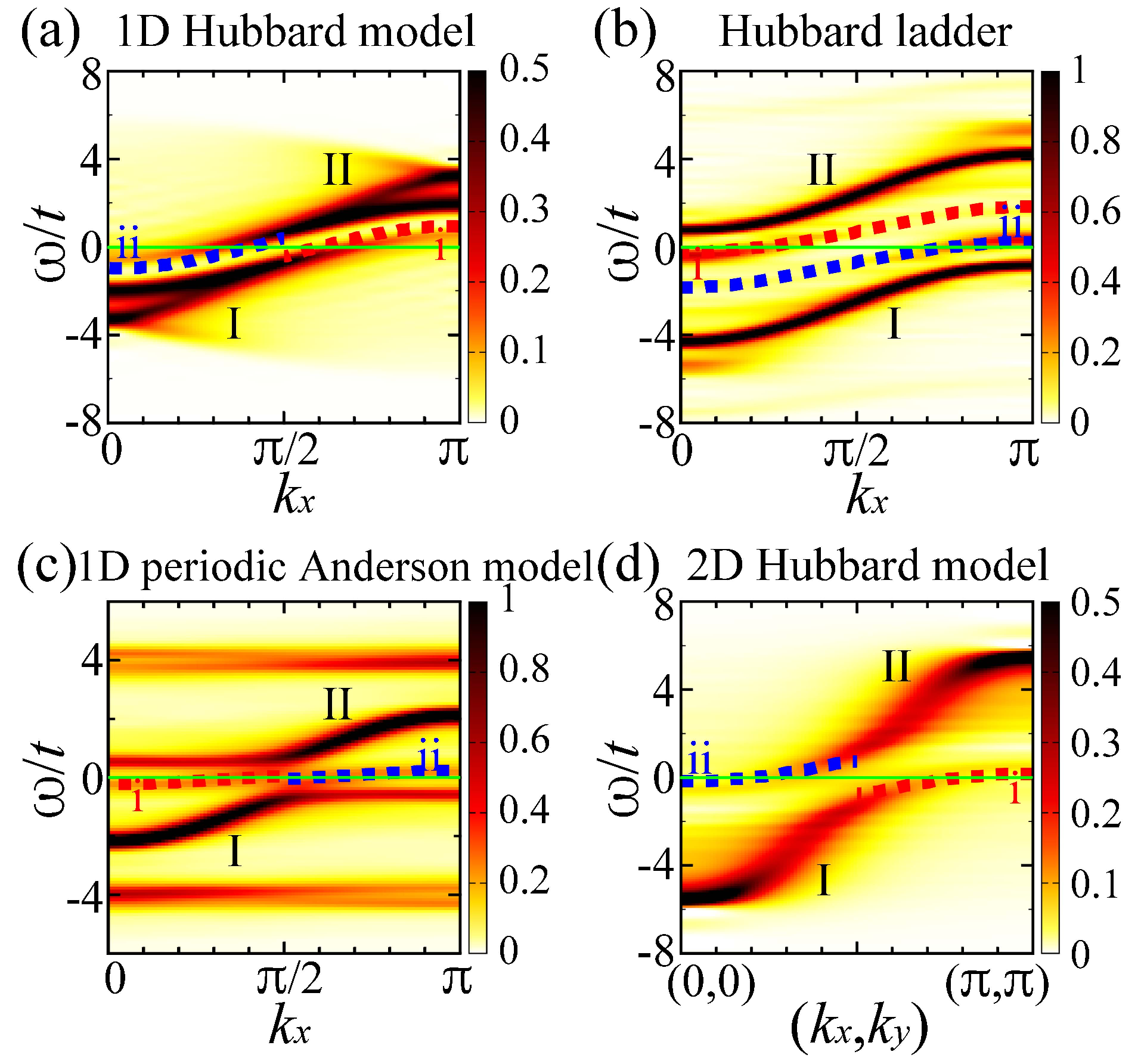}
\caption{Identification of electronic modes in the 1D Hubbard model [(a)], Hubbard ladder [(b)], 1D PAM [(c)], and 2D Hubbard model along $(0,0)$--$(\pi,\pi)$ [(d)]. 
Modes I and II are the dominant lower and upper modes that exist even at $T=0$, respectively. 
Modes i and ii are the temperature-induced modes corresponding to the hole- and  electron-doping-induced states, respectively. 
The dashed red curves and dashed blue curves indicate $\omega=e^{\rm spin}_{{\bm k}-{\bm k}^-_{\rm F}}+\mu_-$ [Eq. (\ref{eq:dispersion1})] 
and $\omega=-e^{\rm spin}_{-{\bm k}+{\bm k}^+_{\rm F}}+\mu_+$ [Eq. (\ref{eq:dispersion2})] 
in the ${\bm k}$ regimes of hole- and electron-doping-induced states, respectively, 
using the dispersion relations of the spin modes ($\omega=e^{\rm spin}_{\bm k}$) extracted from Figs. \ref{fig:Akw}(a-5)--\ref{fig:Akw}(d-5). 
The dominant part of the spin mode was used in (c). The spectral-weight distributions are the same as those in Figs. \ref{fig:Akw}(a-2)--\ref{fig:Akw}(d-2). 
The green lines indicate $\omega=0$.}
\label{fig:AkwEsk}
\end{figure}

As shown in Fig. \ref{fig:AkwEsk}, 
the emergent modes (modes i and ii in Fig. \ref{fig:AkwEsk}) can be interpreted reasonably well as originating from the spin mode, 
essentially exhibiting the spin-mode dispersion relation shifted by ${\bm k}^-_{\rm F}$ and $\mu_-$ 
(top of the lower band) [Eq. (\ref{eq:dispersion1}); the dashed red curves in Fig. \ref{fig:AkwEsk}] 
and the inverted spin-mode dispersion relation shifted by ${\bm k}^+_{\rm F}$ and $\mu_+$ 
(bottom of the upper band) [Eq. (\ref{eq:dispersion2}); the dashed blue curves in Fig. \ref{fig:AkwEsk}] 
in the momentum regimes where hole- and electron-doping-induced states appear, 
respectively (Sec. \ref{sec:regimes}). 
\par
As the temperature increases, the spectral weights of the emergent modes increase 
[Figs. \ref{fig:Akw}(a-1)--\ref{fig:Akw}(a-3), \ref{fig:Akw}(b-1)--\ref{fig:Akw}(b-3), \ref{fig:Akw}(c-1)--\ref{fig:Akw}(c-3), and \ref{fig:Akw}(d-1)--\ref{fig:Akw}(d-3)] 
because the Boltzmann weights of the spin excited states and doped states increase 
[Eq. (\ref{eq:AkwT2}); Figs. \ref{fig:AneU034} and \ref{fig:AS1D}; Secs. \ref{sec:dopedStates} and \ref{sec:reason}]. 
If the bandwidth of the spin excitation is comparable to the electronic band gap, 
the emergent modes can cross the Fermi level ($\omega=0$) 
and gain considerable spectral weight, forming a band structure 
that can be regarded as metallic in the high-temperature regime [Figs. \ref{fig:Akw}(a-1)--\ref{fig:Akw}(d-1)]. 
\par
In the 1D PAM, because the dispersion relation of the spin mode is almost flat [Fig. \ref{fig:Akw}(c-5)], 
the emergent electronic modes also exhibit almost flat dispersion relations, 
which are located almost on the Fermi level ($\omega=0$) and cross the Fermi level [modes i and ii in Fig. \ref{fig:AkwEsk}(c)]. 
A possible relation of the emergent modes to the metallic behavior observed in Kondo insulators 
\cite{SmB6dHvAScience,SmB6dHvANatPhys,SmB6dHvAiScience,YbB12dHvAJPhysCM,YbB12SdHScience,YbB12heatTransNatPhys,YbB12SdHNatPhys,YbB12HallAnomaly,YbB12SdHQM} has been discussed in Ref. \cite{KohnoKLM}. 
\subsection{Remarks on effective emergent modes} % ------ Remarks on effective emergent modes ------
\label{sec:effectiveModes}
In the high-temperature regime, various processes can substantially contribute to the spectral function. 
The dispersion relations of effective (dominant) emergent modes representing accumulated emergent modes 
from all possible processes can deviate from Eqs. (\ref{eq:dispersion1}) and (\ref{eq:dispersion2}). 
For example, contributions from one-hole-doped (one-electron-doped) excited states, 
whose energies are higher than the one-hole-doped (one-electron-doped) ground-state energy, can effectively increase 
$|\mu_-|$ in Eq. (\ref{eq:dispersion1}) ($\mu_+$ in Eq. (\ref{eq:dispersion2})). 
In addition, because spin excited states in doped systems and larger-spin states can have excitation energies 
different from $e^{\rm spin}_{\bm k}$, the effective dispersion relation can deviate 
from Eqs. (\ref{eq:dispersion1}) and (\ref{eq:dispersion2}). 
Hence, depending on the relevant excitations in the temperature regime 
and their energies and values of the matrix elements in the spectral function, 
the dispersion relations of the effective emergent modes can deviate from Eqs. (\ref{eq:dispersion1}) and (\ref{eq:dispersion2}). 
\par
Accordingly, the conditions under which the effective (dominant) emergent modes cross the Fermi level can deviate 
from Eqs. (\ref{eq:conditions}) and (\ref{eq:conditionGapless}) depending on the temperature. 
The conditions for the insulator-metal crossover given by Eqs. (\ref{eq:conditions}) and (\ref{eq:conditionGapless}) are 
the conditions under which the emergent mode due to the processes 
between the spin excited states of Mott and Kondo insulators 
and the one-hole-doped or one-electron-doped ground state [Fig. \ref{fig:ExcitationCartoon}(b)] crosses the Fermi level, 
which generally ensure the existence of spectral weights along the dispersion relation of this mode 
that crosses the Fermi level for $T>0$ as long as the matrix elements of these processes are nonzero [Eq. (\ref{eq:AkwT2})]. 
\par
\section{Comparisons with previous studies} % ------ Comparisons with previous studies ------
\subsection{Comparisons with antiferromagnetic mean-field approximation} % ------ Comparisons with antiferromagnetic mean-field approximation ------
\begin{figure}
\includegraphics[width=\linewidth]{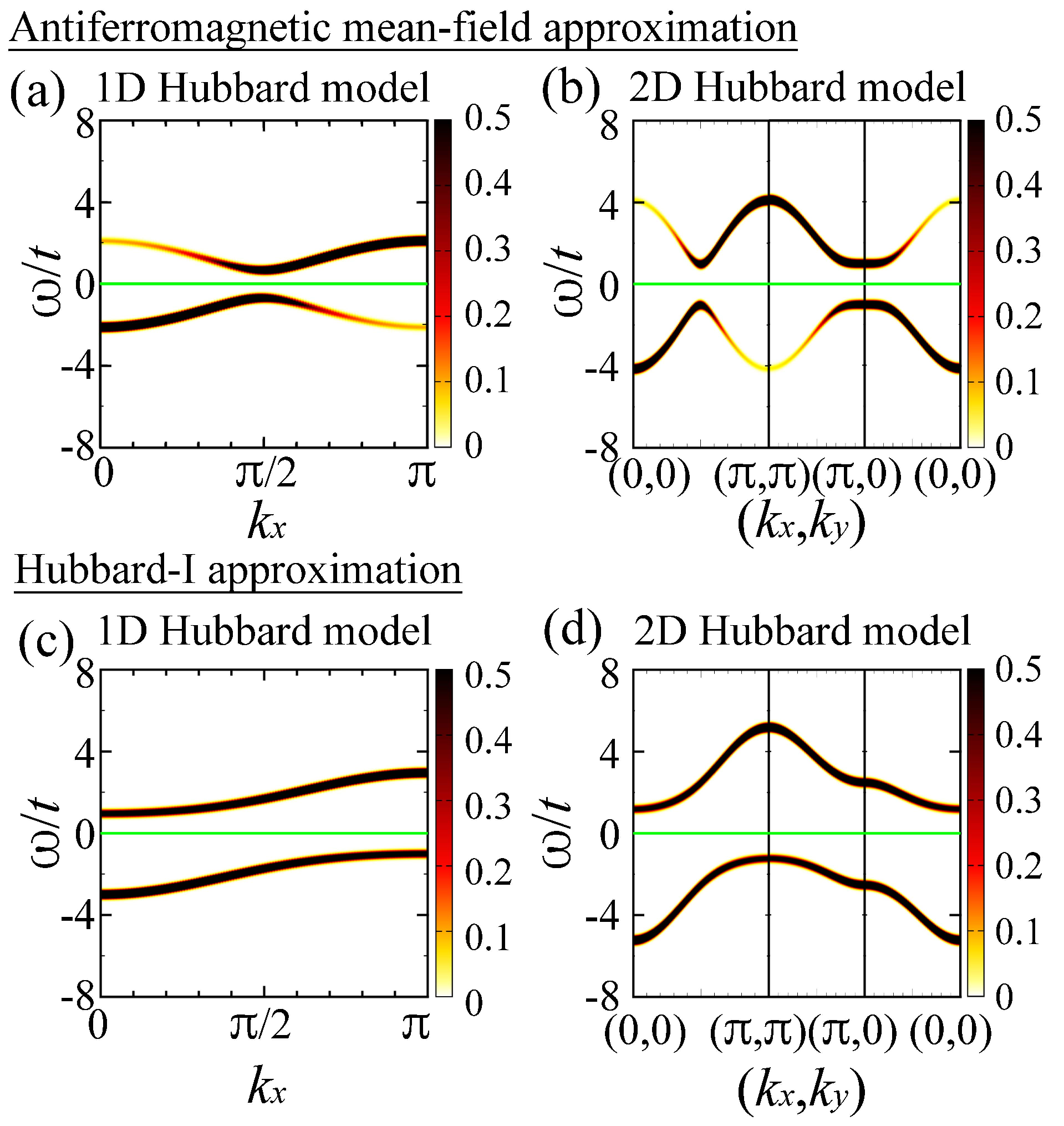}
\caption{(a), (b) $A^c({\bm k},\omega)t$ obtained using the antiferromagnetic mean-field approximation 
in the 1D Hubbard model for $U/t=3.4$ and $m_{\rm s}=0.2$ [(a)] and 2D Hubbard model for $U/t=5$ and $m_{\rm s}=0.2$ [(b)].
(c), (d) $A^c({\bm k},\omega)t$ obtained using the Hubbard-I approximation 
in the 1D Hubbard model for $U/t=3.4$ [(c)] and 2D Hubbard model for $U/t=5$ [(d)]. 
The green lines indicate $\omega=0$. Gaussian broadening with a standard deviation of $0.1t$ was used.}
\label{fig:AkwMF}
\end{figure}

\label{sec:AFMF}
In the antiferromagnetic mean-field approximation \cite{SlaterAF, PennAF}, the spectral function can be obtained as 
\begin{equation}
\label{eq:AkwMF}
A^c({\bm k},\omega)=W^+_{\bm k}\delta(\omega-E^+_{\bm k})+W^-_{\bm k}\delta(\omega-E^-_{\bm k}),
\end{equation}
where 
\begin{align}
\label{eq:EkMF}
W^{\pm}_{\bm k}&=\frac{1}{2}\left(1+\frac{\varepsilon_{\bm k}}{E^{\pm}_{\bm k}}\right),
\quad E^{\pm}_{\bm k}=\pm\sqrt{\varepsilon_{\bm k}^2+m_{\rm s}^2U^2},\nonumber\\
\varepsilon_{\bm k}&=-2t\sum_{i=1}^d \cos k_i
\end{align}
for the Hubbard model at half-filling on a $d$-dimensional cubic lattice ($k_1=k_x$, $k_2=k_y$, and $k_3=k_z$). 
The sublattice magnetization per site is denoted as $m_{\rm s}$. 
As the temperature changes, only the value of $m_{\rm s}$ changes in Eqs. (\ref{eq:AkwMF}) and (\ref{eq:EkMF}). 
Above the antiferromagnetic transition temperature ($m_{\rm s}=0$), 
the spectral function is exactly the same as that of the noninteracting case ($U=0$). 
Below the transition temperature, because an energy gap opens around the Fermi level, 
the electronic modes do not cross the Fermi level, as shown in Figs. \ref{fig:AkwMF}(a) and \ref{fig:AkwMF}(b); 
the temperature-driven insulator-metal crossover [Figs. \ref{fig:Akw}(a-1)--\ref{fig:Akw}(a-3) and \ref{fig:Akw}(d-1)--\ref{fig:Akw}(d-3); Sec. \ref{sec:conditions}] does not occur 
in the antiferromagnetic mean-field approximation. 
\par
The momenta at the bottom of the upper band and top of the lower band ${\bm k}^{\pm}_{\rm F}=\frac{\bm \pi}{2}$ 
are consistent with those of the 1D and 2D Hubbard models at zero temperature 
[Figs. \ref{fig:Akw}(a-4), \ref{fig:Akw}(d-4), and \ref{fig:AS1D}(b)] 
\cite{KohnoRPP,NoceraDMRT_T,GroberQMC,PreussQP,Kohno2DHub,Preuss1DHub,CPTPRL,CPTPRB} 
but differ from those at nonzero temperatures [Figs. 1(a-1)--1(a-3), 1(d-1)--1(d-3), and \ref{fig:AS1D}(a)] \cite{TPQSLanczos,PreussQP,GroberQMC,KuzminCPT,VCALanczos,NoceraDMRT_T}. 
\subsection{Comparisons with Hubbard-I approximation} % ------ Comparisons with Hubbard-I approximation ------
\label{sec:HubI}
In the Hubbard-I approximation \cite{HubbardI}, the spectral function can be obtained as 
\begin{equation}
\label{eq:AkwHubI}
A^c({\bm k},\omega)=\frac{\omega}{{\tilde E}^+_{\bm k}-{\tilde E}^-_{\bm k}}
\left[\delta(\omega-{\tilde E}^+_{\bm k})-\delta(\omega-{\tilde E}^-_{\bm k})\right],
\end{equation}
where 
\begin{equation}
\label{eq:EkHubI}
{\tilde E}^{\pm}_{\bm k}=\frac{\varepsilon_{\bm k}\pm\sqrt{\varepsilon_{\bm k}^2+U^2}}{2}
\end{equation}
for the Hubbard model at half-filling on a $d$-dimensional cubic lattice. For the Hubbard ladder, 
\begin{equation}
\varepsilon_{(k_x,0)}=-2t\cos k_x-t_\perp,\quad\varepsilon_{(k_x,\pi)}=-2t\cos k_x+t_\perp.
\end{equation}
In this approximation, the spectral function does not change with temperature. 
Two electronic excited states exist at each momentum for $U>0$ in the Hubbard model: 
one is in the $\omega>0$ regime and the other is in the $\omega<0$ regime, 
because $|\varepsilon_{\bm k}|<\sqrt{\varepsilon_{\bm k}^2+U^2}$ in Eq. (\ref{eq:EkHubI}). 
\par
\begin{figure}
\includegraphics[width=\linewidth]{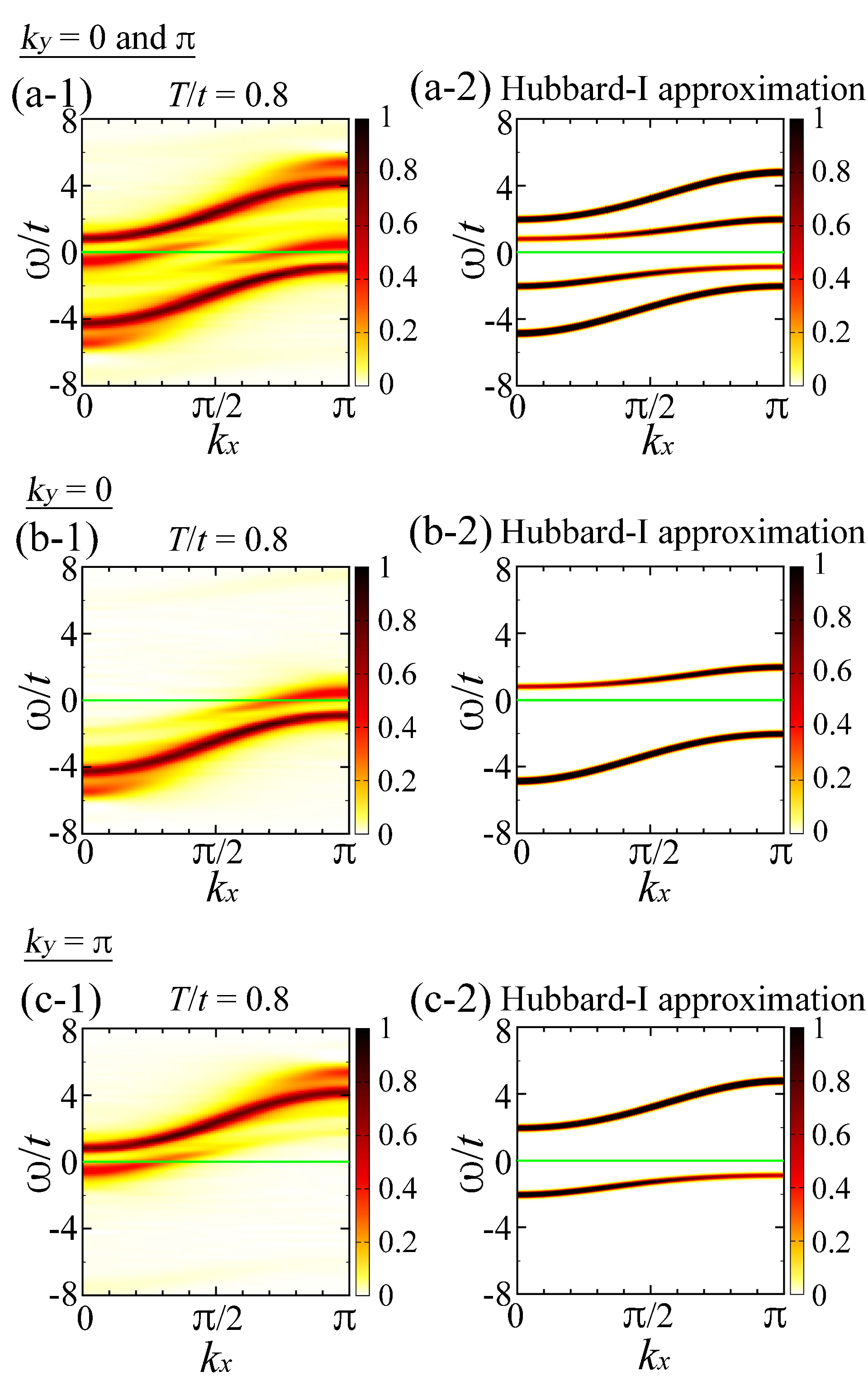}
\caption{Comparisons with the Hubbard-I approximation in the Hubbard ladder for $U/t=4$ and $t_{\perp}/t=2$. 
(a-1)--(c-1) $A^c(k_x,0,\omega)t+A^c(k_x,\pi,\omega)t$ [(a-1)], $A^c(k_x,0,\omega)t$ [(b-1)], and $A^c(k_x,\pi,\omega)t$ [(c-1)] 
at $T/t=0.8$. 
(a-2)--(c-2) $A^c(k_x,0,\omega)t+A^c(k_x,\pi,\omega)t$ [(a-2)], $A^c(k_x,0,\omega)t$ [(b-2)], and $A^c(k_x,\pi,\omega)t$ [(c-2)] 
obtained using the Hubbard-I approximation. 
The green lines indicate $\omega=0$. Gaussian broadening with a standard deviation of $0.1t$ was used.}
\label{fig:HubILad}
\end{figure}

In the 1D and 2D Hubbard models, the bottom of the upper band and top of the lower band are located 
at ${\bm k}={\bm 0}$ and ${\bm \pi}$, respectively, in the Hubbard-I approximation 
[Figs. \ref{fig:AkwMF}(c) and \ref{fig:AkwMF}(d)], 
which differ from those of the 1D and 2D Hubbard models at zero temperature [Figs. \ref{fig:Akw}(a-4), \ref{fig:Akw}(d-4), and \ref{fig:AS1D}(b)] 
\cite{KohnoRPP,NoceraDMRT_T,GroberQMC,PreussQP,Kohno2DHub,Preuss1DHub,CPTPRL,CPTPRB} 
but are consistent with those at nonzero temperatures [Figs. 1(a-1)--1(a-3), 1(d-1)--1(d-3), and \ref{fig:AS1D}(a)] 
\cite{TPQSLanczos,PreussQP,GroberQMC,KuzminCPT,VCALanczos,NoceraDMRT_T}. 
The electronic modes do not cross the Fermi level for $U>0$ in the Hubbard model 
[Eqs. (\ref{eq:AkwHubI}) and (\ref{eq:EkHubI})], 
as shown in Figs. \ref{fig:AkwMF}(c), \ref{fig:AkwMF}(d), and \ref{fig:HubILad}(a-2). 
Additionally, the spectral function does not change with temperature. 
Hence, the temperature-driven insulator-metal crossover 
[Figs. \ref{fig:Akw}(a-1)--\ref{fig:Akw}(a-3), \ref{fig:Akw}(b-1)--\ref{fig:Akw}(b-3), and \ref{fig:Akw}(d-1)--\ref{fig:Akw}(d-3); Sec. \ref{sec:conditions}] 
does not occur in the Hubbard-I approximation. 
\par
The overall spectral features obtained using the Hubbard-I approximation in Hubbard models 
[Figs. \ref{fig:AkwMF}(c), \ref{fig:AkwMF}(d), and \ref{fig:HubILad}(a-2)] are similar to those of the numerical results 
in the high-temperature regime [Figs. \ref{fig:Akw}(a-1), \ref{fig:Akw}(b-1), \ref{fig:Akw}(d-1), and \ref{fig:HubILad}(a-1)]. 
However, even in the high-temperature regime, the $\omega$ levels and $k_y$ values of the in-gap modes 
in the Hubbard ladder are not correctly described in the Hubbard-I approximation. 
If the modes in Figs. \ref{fig:HubILad}(a-1) and \ref{fig:HubILad}(a-2) are identified according to the $\omega$ at each $k_x$, 
the $k_y$ values of the in-gap modes in the Hubbard-I approximation [Figs. \ref{fig:HubILad}(b-2) and \ref{fig:HubILad}(c-2)] 
differ from those of the numerical results [Figs. \ref{fig:HubILad}(b-1) and \ref{fig:HubILad}(c-1)]. 
\par
To identify the emergent modes with respect to $k_y$, we consider the Hubbard ladder 
in the large-$t_\perp/t$ and large-$U/t$ regime \cite{KohnoHubLadder}. 
The ground state can be effectively described as the direct-product state of the spin-singlet rungs, which has $k_y=0$: 
$|{\rm GS}\rangle_N^{0,{\bm 0}}$. 
The spin excited state is obtained by replacing one of the spin-singlet rungs with a spin-triplet rung, 
which has $k_y=\pi$ \cite{KohnoHubLadder}: $|{\rm Spin}\rangle_N^{s^z_{\rm T},(q_x,\pi)}$. 
The one-hole-doped ground state is obtained by removing an electron from the top of the lower band 
[Fig. \ref{fig:Akw}(b-4)], which has $k_y=0$ [Fig. \ref{fig:HubILad}(b-1)]: $|{\rm GS}\rangle_{N-1}^{\zeta,(-\pi,0)}$. 
The one-electron-doped ground state is obtained by adding an electron to the bottom of the upper band 
[Fig. \ref{fig:Akw}(b-4)], which has $k_y=\pi$ [Fig. \ref{fig:HubILad}(c-1)]: $|{\rm GS}\rangle_{N+1}^{\zeta,(0,\pi)}$ 
\cite{KohnoHubLadder}. 
The matrix elements 
$_{N-1}^{\zeta,(-\pi,0)}\langle{\rm GS}|c_{(k_x,\pi),\sigma}|{\rm Spin}\rangle_N^{s^z+\zeta,(k_x-\pi,\pi)}$ and 
$_{N+1}^{\zeta,(0,\pi)}\langle{\rm GS}|c_{(k_x,0),\sigma}^{\dagger}|{\rm Spin}\rangle_N^{-s^z+\zeta,(-k_x,\pi)}$ can be nonzero. 
Thus, the electronic mode due to the spin excited state $|{\rm Spin}\rangle_N^{s^z_{\rm T},(k_x-\pi,\pi)}$ emerges 
in $A^c(k_x,\pi,\omega)$, exhibiting $\omega={\bar e}^{\rm spin}_{k_x-\pi}+\mu_-$ [Eq.(\ref{eq:dispersion1})], and 
that of $|{\rm Spin}\rangle_N^{s^z_{\rm T},(-k_x,\pi)}$ emerges in $A^c(k_x,0,\omega)$, 
exhibiting $\omega=-{\bar e}^{\rm spin}_{-k_x}+\mu_+$ [Eq.(\ref{eq:dispersion2})], where 
${\bar e}^{\rm spin}_{k_x}$ denotes the spin-excitation energy as a function of $k_x$ at $k_y=\pi$ [Fig. \ref{fig:Akw}(b-5)]. 
\par
As shown in Figs. \ref{fig:HubILad}(b-1) and \ref{fig:HubILad}(c-1), the electronic modes emerge at the same $k_y$ values in the spectral function 
as expected from the spin-excitation origin described above. 
In contrast, in the Hubbard-I approximation, the upper (lower) mode near $\omega=0$ [Fig. \ref{fig:HubILad}(a-2)] 
whose dispersion relation is close to that of the upper (lower) emergent mode 
[Fig. \ref{fig:HubILad}(a-1)] has $k_y$ different from that of the upper (lower) emergent mode 
[Figs. \ref{fig:HubILad}(b-1), \ref{fig:HubILad}(b-2), \ref{fig:HubILad}(c-1), and \ref{fig:HubILad}(c-2)]. 
If the in-gap modes are identified according to the $k_y$ values, 
the in-gap mode at $k_y=0$ is lower in $\omega$ than that at $k_y=\pi$ at each $k_x$ [Figs. \ref{fig:HubILad}(a-1)--\ref{fig:HubILad}(c-1)]; 
the results obtained using the Hubbard-I approximation are opposite to this [Figs. \ref{fig:HubILad}(a-2)--\ref{fig:HubILad}(c-2)]. 
\par
\begin{figure}
\includegraphics[width=\linewidth]{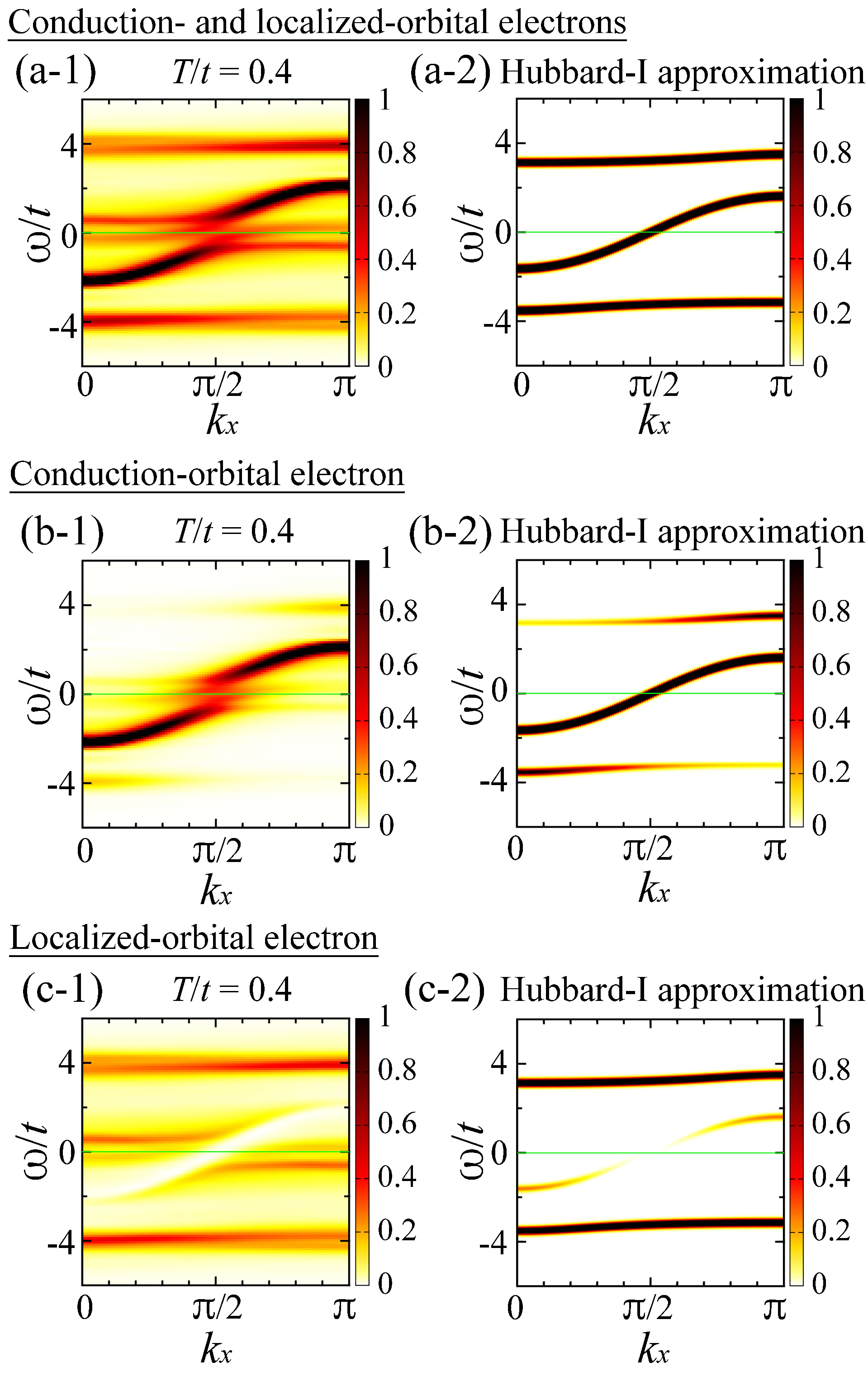}
\caption{Comparisons with the Hubbard-I approximation in the 1D PAM for $U/t=6$ and $t_{\rm K}/t=1.2$. 
(a-1)--(c-1) $A^c(k_x,\omega)t+A^f(k_x,\omega)t$ [(a-1)], $A^c(k_x,\omega)t$ [(b-1)], and $A^f(k_x,\omega)t$ [(c-1)] at $T/t=0.4$. 
(a-2)--(c-2) $A^c(k_x,\omega)t+A^f(k_x,\omega)t$ [(a-2)], $A^c(k_x,\omega)t$ [(b-2)], and $A^f(k_x,\omega)t$ [(c-2)] 
obtained using the Hubbard-I approximation. 
The green lines indicate $\omega=0$. Gaussian broadening with a standard deviation of $0.1t$ was used.}
\label{fig:HubIPAM}
\end{figure}

In the PAM, the Green functions for $U=2\Delta$ are obtained using the Hubbard-I approximation as follows: 
\begin{equation}
\label{eq:HubIPAM}
\begin{split}
G^c_{{\bm k},\sigma}(z)&=\frac{(z+\Delta)(z-\Delta)}{(z-\varepsilon_{\bm k})(z+\Delta)(z-\Delta)-t_{\rm K}^2z},\\
G^f_{{\bm k},\sigma}(z)&=\frac{(z-\epsilon_{\bm k})z}{(z-\varepsilon_{\bm k})(z+\Delta)(z-\Delta)-t_{\rm K}^2z},
\end{split}
\end{equation}
where $z=\omega+i\epsilon$ for $\epsilon\rightarrow +0$ \cite{HubbardIPAM}. There are three poles for $U>0$: 
one corresponds to a gapless dispersing mode primarily due to the conduction-orbital electron, 
and the other two correspond to almost flat modes at high $|\omega|$ primarily due to the localized-orbital electron 
[Figs. \ref{fig:HubIPAM}(a-2)--\ref{fig:HubIPAM}(c-2)]. 
\par
The overall spectral features of the Hubbard-I approximation [Fig. \ref{fig:HubIPAM}(a-2)] are consistent with 
those of the numerical results in the high-temperature regime [Figs. \ref{fig:Akw}(c-1) and \ref{fig:HubIPAM}(a-1)]. 
However, the hybridization around $\omega=0$ [disconnection of the dispersing mode of the conduction-orbital electron 
at $k_x=\frac{\pi}{2}$; almost flat dispersion relations around $k_x=0$ and $\pi$ near $\omega=0$ 
in the excitation of the localized-orbital electron], 
which occurs even at zero temperature [Fig. \ref{fig:Akw}(c-4)], is not properly described in the Hubbard-I approximation 
[Figs. \ref{fig:HubIPAM}(a-2)--\ref{fig:HubIPAM}(c-2)]. 
In addition, the temperature-induced modes exhibiting almost flat dispersion relations almost on the Fermi level ($\omega=0$) 
around $k_x=0$ and $\pi$ primarily in the excitation of the localized-orbital electron 
[Figs. \ref{fig:HubIPAM}(a-1) and \ref{fig:HubIPAM}(c-1); modes i and ii in Fig. \ref{fig:AkwEsk}(c)] 
are not properly described in the Hubbard-I approximation [Figs. \ref{fig:HubIPAM}(a-2) and \ref{fig:HubIPAM}(c-2)]. 
\subsection{Comparisons with previous numerical studies} % ------ Comparisons with previous numerical studies ------
\label{sec:comp}
Numerical calculations have indicated that the band structures change with temperature in the 1D and 2D Hubbard models and 1D PAM 
\cite{TPQSLanczos,PreussQP,GroberQMC,KuzminCPT,VCALanczos,NoceraDMRT_T,Matsueda_QMC,PAMMPS}: 
the band structures in the high-temperature regime are similar to those of the Hubbard-I approximation (Sec. \ref{sec:HubI}), 
whereas those in the low-temperature regime are similar to those of the antiferromagnetic mean-field approximation (Sec. \ref{sec:AFMF}) 
for the Hubbard models. 
These features have been interpreted primarily in terms of spin correlations and dressed electronic quasiparticles 
for the 2D Hubbard model and 1D quasiparticles such as spinons and (anti)holons for the 1D Hubbard model. 
\par
For the 2D Hubbard model, in Refs. \cite{GroberQMC,PreussQP}, 
the reason why the band structure in the low-temperature regime is similar to that of the antiferromagnetic mean-field approximation 
has been considered that the spin correlation length becomes comparable to the cluster size of the Monte Carlo simulation; 
the system can be effectively regarded as antiferromagnetically ordered. 
In contrast, in the high-temperature regime, 
because the spin correlation length becomes shorter, 
the band structure has been considered similar to that of the Hubbard-I approximation 
in which the spatial spin correlation is neglected. 
Thus, the Hubbard-I-like band structure has been considered to be primarily attributed to the reduction in the spin correlation 
through thermal fluctuations, in contrast to the interpretation presented in this paper, i.e., 
that the spin excited states involved in the thermal states [Fig. \ref{fig:ExcitationCartoon}(b)] are the main cause of the Hubbard-I-like band structure. 
The reason why the spectral weights within the band gap ($\mu_-<\omega<\mu_+$) fade away 
instead of the continuous deformation of the dispersion relation from the Hubbard-I-like band structure 
into the low-temperature band structure has not been considered. 
\par
The behavior of the modes responsible for the Hubbard-I-like band structure in the low-temperature regime is 
as nontrivial as that of the zero-temperature doping-induced states. In the doping-driven Mott transition, 
how the free-electron-like mode around the Fermi level in the large-doping regime changes toward the Mott transition 
(whether the dispersion relation becomes flat as expected for the Fermi liquid quasiparticle \cite{LandauFL} 
with a divergent effective mass \cite{BrinkmanRice} or not; 
if not, how the mode behaves and why) has been a central question 
\cite{ImadaRMP,SakaiImadaPRL,SakaiImadaPRB,ImadaCofermionPRL,ImadaCofermionPRB,PhillipsRMP,PhillipsRPP,EderOhtaIPES,EderOhta2DHub,q1DCPTDMRG,KohnoRPP,Kohno1DHub,Kohno2DHub,Kohno1DtJ,Kohno2DtJ,KohnoDIS,KohnoAF,KohnoSpin,KohnoHubLadder,KohnoGW,KohnoKLM}. 
It has been shown that the free-electron-like mode above (below) the Fermi level in a hole-doped (electron-doped) system 
loses spectral weight toward the Mott transition, exhibiting a momentum-shifted magnetic dispersion relation, 
because the doping-induced states can be essentially identified as the spin excited states of Mott and Kondo insulators 
\cite{KohnoRPP,Kohno1DHub,Kohno2DHub,Kohno1DtJ,Kohno2DtJ,KohnoDIS,KohnoAF,KohnoSpin,KohnoHubLadder,KohnoGW,KohnoKLM}. 
Regarding the temperature-driven change in the band structure, it is shown in this paper 
that the modes induced by temperature are primarily due to the spin excited states 
involved in the thermal state [Sec. \ref{sec:emergentModes}; Fig. \ref{fig:ExcitationCartoon}(b)]. 
Because their contributions to the thermal state decrease to zero as the temperature is lowered, 
the spectral weights of the emergent modes fade away. 
In addition, the dispersion relations of the emergent modes in the low-temperature regime can be effectively 
expressed as momentum-shifted magnetic dispersion relations from the band edges, 
reflecting the spin excited states. 
These characteristics are continuously connected to the doping-driven Mott transition 
\cite{KohnoRPP,Kohno1DHub,Kohno2DHub,Kohno1DtJ,Kohno2DtJ,KohnoDIS,KohnoAF,KohnoSpin,KohnoHubLadder,KohnoGW,KohnoKLM} (Sec. \ref{sec:dopedStates}). 
\par
For the 1D Hubbard model, in Ref. \cite{NoceraDMRT_T}, it has been recognized that 
the origin of the emergent modes can be traced back to the zero-temperature doping-induced states \cite{Kohno1DHub}, 
and the temperature evolution has been interpreted in terms of 1D quasiparticles 
such as spinons and (anti)holons for electronic excitation \cite{1dtJT,UinftyT} 
[spin-charge separation characteristic of 1D electronic excitation (Sec. \ref{sec:spinChargeSeparation})]. 
The effective hopping (bandwidth) at high temperatures has been considered to be determined 
by the holon excitation that can be described as a spinless quasiparticle with hopping $t$, 
and the band structure has been interpreted using the Hubbard-I approximation with effective $U$. 
\par
In this paper, the emergent modes are interpreted from a broader perspective 
in terms of the spin-charge separation of Mott and Kondo insulators (Sec. \ref{sec:spinChargeSeparation}), 
which does not depend on the spatial dimension. 
The properties of the emergent modes within the band gap ($\mu_-<\omega<\mu_+$) are primarily determined by the spin excited states 
in the thermal state, regardless of whether the spatial dimension is one or larger, 
whether the electronic excitation is described in terms of spinons and (anti)holons or not, 
whether the spin excitation is gapless or not, or 
whether the system is antiferromagnetically ordered or not. 
\par
The filling of the band gap with spectral weights, which is also called the melting of the gap \cite{lightPump,NoceraDMRT_T}, 
has been indicated by numerical calculations at nonzero temperatures for the 1D Hubbard model and 1D PAM in Refs. \cite{NoceraDMRT_T,Matsueda_QMC,PAMMPS}. 
In this paper, it is explicitly demonstrated that the emergent modes reflecting the spin excitation can cross the Fermi level, 
gain considerable spectral weight, and form a robust band structure that can be regarded as metallic 
not only in the 1D Hubbard model and 1D PAM but also in the Hubbard ladder and 2D Hubbard model. 
In addition, the mechanism and conditions for the insulator-metal crossover in Mott and Kondo insulators 
in general are clarified 
in terms of the spin excitation ($e^{\rm spin}_{\bm q}$) and charge excitation ($\mu_{\pm}$) of the Mott and Kondo insulators (Sec. \ref{sec:conditions}). 
\subsection{Spin-charge separation of Mott and Kondo insulators} % ------ Spin-charge separation of Mott and Kondo insulators ------
\label{sec:spinChargeSeparation}
Spin-charge separation means that the lowest spin-excitation energy is different from the lowest charge-excitation energy. 
In 1D interacting metals, the spin velocity $v_{\rm s}$, which is characterized by a spinon, is generally different from 
the charge velocity $v_{\rm c}$, which is characterized by a holon or an antiholon \cite{TakahashiBook,Essler}: 
the lowest spin-excitation energy $\Delta E_{\rm s}=\frac{2\pi}{L}v_{\rm s}$ differs 
from the lowest charge-excitation energy $\Delta E_{\rm c}=\frac{2\pi}{L}v_{\rm c}$, 
where $L$ denotes the number of sites on a chain. 
Thus, spin-charge separation occurs in 1D interacting metals in the energy scale of the order of $\frac{1}{L}$. 
\par
On the other hand, in Mott and Kondo insulators, 
spin-charge separation occurs in any spatial dimensions. 
In the large-$U/t$ Hubbard model, the spin-excitation energies are of the order of $J=\frac{4t^2}{U}$, 
whereas the lowest charge-excitation energy is the charge gap of the order of $U$ \cite{TakahashiBook,Essler}. 
Owing to the spin-charge separation of the Mott and Kondo insulators, the low-energy properties can be described 
using spin models such as the Heisenberg model, by neglecting high-energy charge degrees of freedom. 
\par
The emergent electronic modes at nonzero temperature discussed in this paper, along with the doping-induced states in Refs. \cite{KohnoRPP,Kohno1DHub,Kohno2DHub,Kohno1DtJ,Kohno2DtJ,KohnoDIS,KohnoAF,KohnoSpin,KohnoHubLadder,KohnoGW,KohnoKLM}, 
reflect the spin-charge separation of Mott and Kondo insulators 
(existence of spin excited states with excitation energies lower than the charge gap). 
Thus, the emergence of the electronic states reflecting the spin excitation is a fundamental and general characteristic 
around Mott and Kondo insulators 
regardless of the spatial dimension, antiferromagnetic order, or quasiparticle picture 
\cite{KohnoRPP,Kohno1DHub,Kohno2DHub,Kohno1DtJ,Kohno2DtJ,KohnoDIS,KohnoAF,KohnoSpin,KohnoHubLadder,KohnoGW,KohnoKLM}. 
\par
In a band insulator which is described in terms of noninteracting electrons, spin-charge separation does not occur. 
The spin excitation is obtained as a particle--hole excitation: 
a spin excited state with the $z$ component of spin $S^z=1$ from the ground state with $S^z=0$ is obtained 
by removing a down-spin electron from the lower band and adding an up-spin electron to the upper band. 
The lowest spin-excitation energy is equal to the band gap (lowest charge-excitation energy). 
Thus, electronic states reflecting the spin excitation of the band insulator do not emerge within the band gap even with doping or temperature rising. 
\par
In mean-field approximations where the effective Hamiltonian is described in terms of noninteracting electronic quasiparticles, 
the spin excitation is obtained as a particle--hole excitation, as in the case of a band insulator. 
Thus, in mean-field approximations such as the antiferromagnetic mean-field approximation, 
electronic states reflecting the spin excitation of the insulator do not emerge within the band gap even with doping or temperature rising. 
By taking into account the spin excitation obtained beyond the mean-field approximations, the emergent modes can be explained \cite{KohnoAF}. 
\subsection{Difference from conventional wisdom} % ------ Difference from conventional wisdom ------
\label{sec:differenceConventional}
In previous studies, changes in the band structure with respect to the temperature have been interpreted 
in terms of quasiparticles such as dressed electronic quasiparticles \cite{GroberQMC}, 
spinons, and (anti)holons \cite{NoceraDMRT_T} for electronic excitation. 
Electronic states that emerge within the band gap ($\mu_-<\omega<\mu_+$) 
should be considered according to conventional wisdom, as follows. 
Eigenstates whose excitation energies from $|{\rm GS}\rangle_N$ are lower than $|\mu_{\pm}|$ 
should exist with $N_{\rm e}=N\pm 1$, 
similar to the electronic excited states in the upper ($N_{\rm e}=N+1$) and lower ($N_{\rm e}=N-1$) bands. 
Because they are invisible in the spectral function at zero temperature, their spectral weights should be zero. 
By increasing the temperature, they would gain spectral weight through thermal fluctuations and 
emerge in the spectral function at nonzero temperature. 
\par
However, there is no eigenstate with $N_{\rm e}=N\pm 1$ whose energy is lower than the ground-state energy 
in the subspace of $N_{\rm e}=N\pm1$ according to the definition of the ground state; i.e., 
no eigenstate with $N_{\rm e}=N\pm1$ exists in the energy regime of the band gap. 
One might consider that the doping-induced states in one-hole-doped (one-electron-doped) systems, 
which emerge within the band gap, should exist as eigenstates with $N_{\rm e}=N-1$ ($N+1$), 
similar to the electronic states in the lower (upper) band. 
However, the doping-induced states in one-hole-doped (one-electron-doped) systems have $N$ electrons 
because they are induced by the addition (removal) of an electron. 
They are essentially identified as the spin excited states of Mott and Kondo insulators ($N_{\rm e}=N$), 
which exhibit momentum-shifted magnetic dispersion relations from the band edges in the electronic spectrum with doping 
(Secs. \ref{sec:holeDope} and \ref{sec:electronDope}; Table \ref{tbl:selectionRule}) 
\cite{KohnoRPP,Kohno1DHub,Kohno2DHub,Kohno1DtJ,Kohno2DtJ,KohnoDIS,KohnoAF,KohnoSpin,KohnoHubLadder,KohnoGW,KohnoKLM}. 
\par
One might also consider that the doping-induced states in two-hole-doped (two-electron-doped) systems, 
which emerge within the band gap, should exist as eigenstates with $N_{\rm e}=N-1$ ($N+1$) 
in the energy regime within the band gap. 
However, the doping-induced states in two-hole-doped (two-electron-doped) systems can be identified 
as the spin excited states with $N_{\rm e}=N-1$ ($N+1$) (Sec. \ref{sec:multi}; Table \ref{tbl:selectionRule2}) 
\cite{KohnoRPP,Kohno1DHub,Kohno2DHub,Kohno1DtJ,Kohno2DtJ,KohnoDIS,KohnoAF,KohnoSpin,KohnoHubLadder,KohnoGW,KohnoKLM}, 
whose energies are higher than the ground-state energy in the subspace of $N_{\rm e}=N-1$ ($N+1$). 
Thus, they should appear outside the band gap [$\omega<\mu_-$ ($\omega>\mu_+$)] 
in the excitation from $|{\rm GS}\rangle_N$. 
According to the selection rules between $|{\rm Spin}\rangle_{N\pm 1}$ and $|{\rm GS}\rangle_N$ 
presented in Table \ref{tbl:selectionRule3}, the dispersion relations are 
\begin{align}
\label{eq:dispersion31}
\omega&={\tilde e}^{\rm spin}_{{\bm k}-{\bm k}^+_{\rm F}}+\mu_+\approx e^{\rm spin}_{{\bm k}-{\bm k}^+_{\rm F}}+\mu_+,\\
\label{eq:dispersion32}
\omega&=-{\tilde e}^{\rm spin}_{-{\bm k}+{\bm k}^-_{\rm F}}+\mu_-\approx -e^{\rm spin}_{-{\bm k}+{\bm k}^-_{\rm F}}+\mu_-,
\end{align}
where ${\tilde e}^{\rm spin}_{\bm q}$ denotes the energy of spin excitation with momentum ${\bm q}$ 
in the subspace of $N_{\rm e}=N\pm 1$. 
This excitation is included in the spectral function even at zero temperature, 
which can be identified as the mode known as the spinon mode in 1D systems \cite{Kohno1DHub,DDMRGAkw,TakahashiBook,Essler} and 
the mode originating from the 1D spinon mode in the $(0,0)$--$(\pi,\pi)$ direction in 2D systems \cite{Kohno2DHub,Kohno2DtJ}. 
The selection rules [Table \ref{tbl:selectionRule3}; Eqs. (\ref{eq:dispersion31}) and ({\ref{eq:dispersion32}})] 
provide a simple interpretation of the electronic excitation 
whose energy is close to the spin-excitation energy of the order of $J(=\frac{4t^2}{U})$ in Mott insulators at zero temperature 
\cite{PreussQP,Dagotto1h,Poilblanc1h,DagottoFlatband,LiuHole2DtJ,DagottoRMP,Kohno1DHub,DDMRGAkw,TakahashiBook,Essler,Kohno2DHub,Kohno2DtJ} regardless of the spatial dimension, antiferromagnetic order, or quasiparticle picture. 
The above argument [Table \ref{tbl:selectionRule3}; Eqs. (\ref{eq:dispersion31}) and ({\ref{eq:dispersion32})] can also be 
extended to the electron-removal (electron-addition) excitation in hole-doped (electron-doped) systems, as described in Sec. \ref{sec:multi}. 
\begin{table}
\caption{Selection rules for $\langle m|c^{\dagger}_{{\bm k},\sigma}|n\rangle$ and 
$\langle m|f^{\dagger}_{{\bm k},\sigma}|n\rangle$ between $|{\rm Spin}\rangle_{N\pm 1}$ and $|{\rm GS}\rangle_N$.
The energy, $z$ component of spin, and momentum are shown in parentheses. 
The $\omega$ values at the bottom of the upper band and top of the lower band at zero temperature 
are denoted as $\mu_+$ and $\mu_-$, respectively: 
$\mu_+=E^{\rm GS}_{N+1}-E^{\rm GS}_N$; $\mu_-=E^{\rm GS}_{N}-E^{\rm GS}_{N-1}$. 
The energy of spin excitation with momentum ${\bm q}$ in the subspace of $N_{\rm e}=N\pm 1$ 
is denoted as ${\tilde e}^{\rm spin}_{\bm q}$. 
The momenta of the ground states with $N_{\rm e}=N+1$ and $N-1$ are denoted as 
${\bm k}^+_{\rm F}$ and $-{\bm k}^-_{\rm F}$, respectively.}
\label{tbl:selectionRule3}
\begin{tabular}{ccc}
\hline\hline
$|m\rangle$&$|n\rangle$&$\omega=E_m-E_n$\\\hline
$|{\rm Spin}\rangle_{N+1}$&$|{\rm GS}\rangle_N$&
\multirow{4}{*}{$\omega={\tilde e}^{\rm spin}_{{\bm k}-{\bm k}^+_{\rm F}}+\mu_+$}\\
$\left(\begin{array}{c}
{\tilde e}^{\rm spin}_{{\bm k}-{\bm k}^+_{\rm F}}+E^{\rm GS}_{N+1}\\
s^z\\
{\bm k}
\end{array}\right)$&
$\left(\begin{array}{c}
E^{\rm GS}_N\\
0\\
{\bm 0}
\end{array}\right)$&\\\hline\hline
$|n\rangle$&$|m\rangle$&$\omega=E_m-E_n$\\\hline
$|{\rm Spin}\rangle_{N-1}$&$|{\rm GS}\rangle_N$&
\multirow{4}{*}{$\omega=-{\tilde e}^{\rm spin}_{-{\bm k}+{\bm k}^-_{\rm F}}+\mu_-$}\\
$\left(\begin{array}{c}
{\tilde e}^{\rm spin}_{-{\bm k}+{\bm k}^-_{\rm F}}+E^{\rm GS}_{N-1}\\
-s^z\\
-{\bm k}
\end{array}\right)$&
$\left(\begin{array}{c}
E^{\rm GS}_N\\
0\\
{\bm 0}
\end{array}\right)$&\\\hline\hline
\end{tabular}\\
\end{table}
\par
The fundamental difference between conventional wisdom and the interpretation presented in this paper is that 
the electronic states with $N_{\rm e}=N\pm 1$ are considered to change with temperature in the former, 
whereas the spin excited states of Mott and Kondo insulators, 
which become significant in the thermal state as the temperature increases, 
are considered to change the electronic spectral features in the latter. 
This interpretation explains why the band structure can change with temperature even in the energy regime 
significantly higher than the temperature (Sec. \ref{sec:reason}) and why electronic modes emerge within the band gap and 
exhibit momentum-shifted spin-mode-like dispersion relations from the band edges 
[Secs. \ref{sec:holeDope} and \ref{sec:electronDope}; Table \ref{tbl:selectionRule}; Fig. \ref{fig:ExcitationCartoon}(b)]. 
In addition, it provides a unified view of the doping-driven \cite{KohnoRPP,Kohno1DHub,Kohno2DHub,Kohno1DtJ,Kohno2DtJ,KohnoDIS,KohnoAF,KohnoSpin,KohnoHubLadder,KohnoGW,KohnoKLM} 
and temperature-driven unconventional spectral features around Mott and Kondo insulators 
without depending on the spatial dimension, antiferromagnetic order, or quasiparticle picture. 
The essence of this interpretation is the spin-charge separation 
(existence of spin excited states with excitation energies lower than the charge gap), 
which is a fundamental and general property of Mott and Kondo insulators 
\cite{KohnoRPP,Kohno1DHub,Kohno2DHub,Kohno1DtJ,Kohno2DtJ,KohnoDIS,KohnoAF,KohnoSpin,KohnoHubLadder,KohnoGW,KohnoKLM}. 
\section{Summary} % ------ Summary ------
The mechanism underlying the change in the band structure with respect to the temperature 
in Mott and Kondo insulators was clarified 
by considering the selection rules in the spectral function and using numerical methods. 
Because the spin excited states of Mott and Kondo insulators are involved in the thermal state at nonzero temperature, 
they can appear in the electronic spectral function and exhibit momentum-shifted magnetic dispersion relations 
from the band edges. 
In addition, one-hole-doped and one-electron-doped states in the thermal state induce electronic states 
that exhibit the momentum-shifted magnetic dispersion relations within the band gap, 
which can also be identified as the spin excited states of Mott and Kondo insulators, 
as in the case of the doping-driven Mott transition. 
Hence, the origin of the emergence of electronic modes (change in the band structure) at nonzero temperature 
can be traced back to the spin excited states of Mott and Kondo insulators; 
the essence is the spin-charge separation of Mott and Kondo insulators 
(existence of spin excited states with excitation energies lower than the charge gap). 
This perspective contrasts with the conventional view where thermal effects on electron-added states ($N_{\rm e}=N+1$), 
electron-removed states ($N_{\rm e}=N-1$), and static spin correlations 
are considered to mainly affect the band structure. 
\par
The present perspective can explain 
(1) why the band structure changes even in the $|\omega|$ regime far higher than the temperature, 
(2) why spectral weights emerge in the $\omega$ regime within the band gap, where the excitation energies are lower than 
the lowest electronic excitation energy from the ground state, and 
(3) why the dispersion relations of the emergent modes are similar to the spin-mode dispersion relation 
shifted by ${\bm k}^{\pm}_{\rm F}$ and $\mu_{\pm}$ (${\bm k}$ and $\omega$ at the band edges). 
\par
The emergent modes can cross the Fermi level if the bandwidth of the spin excitation is comparable to the band gap; 
the band structure can be regarded as metallic. 
The mechanism and conditions for this insulator-metal crossover were clarified 
in terms of the spin-charge separation of Mott and Kondo insulators. 
Additionally, the spectral weight below the Fermi level can increase as the chemical potential is lowered, 
reflecting the change in the spectral-weight distribution with respect to the Fermi level in Mott and Kondo insulators. 
These features cannot be explained in the conventional band picture or mean-field approximations. 
\par
The mechanism of the temperature-driven change in the band structure revealed in this paper is general. 
It does not depend on 
(1) the spatial dimension, 
(2) quasiparticle picture such as electronic quasiparticles, spinons, or (anti)holons, 
(3) whether the system is antiferromagnetically ordered or not, or 
(4) whether the spin excitation is gapless or not. 
This is because it reflects the spin-charge separation 
which is a fundamental and general property of Mott and Kondo insulators. 
\par
Because raising the temperature is far easier than doping Mott and Kondo insulators, 
the emergence of electronic states reflecting the spin excitation of Mott and Kondo insulators 
can be experimentally confirmed for various materials that are difficult to dope. 
Furthermore, the change in the band structure with respect to the temperature, including the insulator-metal crossover, 
can be utilized in optical and electronic devices beyond the conventional rigid-band-based electronics. 
\begin{acknowledgments} % ------ Acknowledgments ------
This work was supported by JSPS KAKENHI Grant Number JP22K03477 and World Premier International Research Center Initiative (WPI), MEXT, Japan. 
Numerical calculations were partly performed on the Numerical Materials Simulator at National Institute for Materials Science. 
\end{acknowledgments}
%\bibliography{apssamp}

\end{document}